\documentclass[11pt]{article}
\pdfoutput=1
\usepackage{color}
\usepackage{amsmath,amssymb,amsfonts,dsfont}
\usepackage{tensor, slashed, empheq}
\usepackage[citebordercolor={.8 .8 1},urlbordercolor ={.8 .8 1},pdfstartview=FitH]{hyperref}
\usepackage{graphicx}
\newcommand{\beq}{\begin{equation}}
\newcommand{\eeq}{\end{equation}}
\newcommand{\beqn}{\begin{eqnarray}}
\newcommand{\eeqn}{\end{eqnarray}}
\newcommand{\pa}{\partial}
\newcommand{\pasls}{\pa\kern-.4em /}
\newcommand{\pasl}{\pa\kern-.55em /}
\newcommand{\Dsls}{D\kern-.52em /}
\newcommand{\Dsl}{D\kern-.65em /}
\newcommand{\qsl}{q\kern-.5em /}
\newcommand{\ksl}{k\kern-.5em /}
\newcommand{\psl}{p\kern-.45em /}
\newcommand{\bAsl}{{\rm A}\kern-.55em /}
\newcommand{\Asl}{A\kern-.55em /}
\newcommand{\Bsl}{B\kern-.55em /}
\newcommand{\epssl}{\epsilon\kern-.45em /}
\newcommand{\Fsl}{F\kern-.65em /\kern.1em}
\newcommand{\cFsl}{{\cal F}\kern-.65em /\kern.2em}
\newcommand{\Gsl}{G\kern-.65em /\kern.2em}
\newcommand{\Jsl}{J\kern-.65em /\kern.2em}
\newcommand{\Psl}{P\kern-.6em /\kern.2em}
\newcommand{\Qsl}{Q\kern-.65em /\kern.2em}
\newcommand{\Hsl}{H\kern-.65em /\kern.2em}
\newcommand{\omsl}{\omega\kern-.65em /\kern.2em}
\newcommand{\Omsl}{\Omega\kern-.65em /\kern.2em}
\newcommand{\Ombsl}{\overline{\Omega}\kern-.65em /\kern.2em}
\newcommand{\cAsl}{{\cal A}\kern-.55em /\kern.2em}
\newcommand{\cDsl}{{\cal D}\kern-.55em /\kern.2em}

\newcommand{\conj}{{\bf c}}
\newcommand{\Sigsl}{\Sigma\kern-.65em /\kern.2em}
\newcommand{\nablasl}{\nabla\kern-.7em /\kern.2em}

\newcommand{\sfLam}{\mathsf{\Lambda}}

\newcommand{\f}{{\kern.5pt{\rm f}}}

\newcommand{\fudge}{}

\numberwithin{equation}{section}

\newcommand{\cg}[3]{\mathcal{G}\indices{_{#1}^{#2}_{#3}}}

\newcommand{\cgsl}[1] {\slashed{\mathcal{G}}_{#1}}
\newcommand{\cgtsl}[1] {\tilde{\slashed{\mathcal{G}}}_{#1}}

\newcommand{\cnm}[3]{\mathcal{N}^{(-)#2}_{#1#3}}
\newcommand{\cnpm}[3]{\mathcal{N}^{(\pm) #2}_{#1#3}}

\newcommand{\cntp}[3]{\tilde{\mathcal{N}}^{(+)#2}_{#1#3}}

\newcommand{\cntpm}[3]{\tilde{\mathcal{N}}^{(\pm) #2}_{#1#3}}

\newcommand{\cnpmsl}[1] {\slashed{\mathcal{N}}^{(\pm)}_{#1}}

\newcommand{\cntpmsl}[1] {\tilde{\slashed{\mathcal{N}}}^{(\pm)}_{#1}}

\begin{document}
\begin{center}

\vspace{1cm} { \LARGE {\bf Fermions and Type IIB Supergravity On Squashed Sasaki-Einstein Manifolds}}

\vspace{1.1cm}
Ibrahima Bah$^{1}$, Alberto Faraggi$^{1}$, Juan I. Jottar$^{2}$ and Robert G. Leigh$^{2}$

\vspace{0.7cm}

{\it $^{1}$Michigan Center for Theoretical Physics, Randall Laboratory of Physics, \\
     University of Michigan, Ann Arbor, MI 48109, U.S.A.}

\vspace{0.7cm}

{\it $^{2}$Department of Physics, University of Illinois,\\
      1110 W. Green Street, Urbana, IL 61801, U.S.A. }

\vspace{0.7cm}

{\tt ibbah@umich.edu, faraggi@umich.edu, jjottar2@illinois.edu, rgleigh@illinois.edu} \\

\vspace{1.5cm}

\end{center}

\begin{abstract}
\noindent
We discuss the dimensional reduction of fermionic modes in a recently found class of consistent truncations of type IIB supergravity compactified on squashed five-dimensional Sasaki-Einstein manifolds. We derive the lower dimensional equations of motion and effective action, and comment on the supersymmetry of the resulting theory, which is consistent with $N=4$ gauged supergravity in $d=5$, coupled to two vector multiplets. We compute fermion masses by linearizing around two $AdS_{5}$ vacua of the theory: one that breaks $N=4$ down to $N=2$ spontaneously, and a second one which preserves no supersymmetries. The truncations under consideration are noteworthy in that they retain massive modes which are charged under a $U(1)$ subgroup of the $R$-symmetry, a feature that makes them interesting for applications to condensed matter phenomena via gauge/gravity duality. In this light, as an application of our general results we exhibit the coupling of the fermions to the type IIB holographic superconductor, and find a consistent further truncation of the fermion sector that retains a single spin-1/2 mode. 

\vspace{30pt}
\end{abstract}

\pagebreak

\tableofcontents

\setcounter{page}{1}
\setcounter{equation}{0}

\section{Introduction}
Recently, consistent truncations of type IIB and 11-$d$ supergravity including massive (charged) modes have sparked a great deal of interest. The relevance of these reductions is two-fold: not only are they novel from the supergravity perspective, but they also constitute an interesting arena to test and extend the ideas of gauge/gravity duality. Indeed, these truncations provide a powerful way of generating solutions of the ten and eleven-dimensional supergravity theories via uplifting of lower dimensional solutions. By definition, this possibility is guaranteed by the consistency of the reduction. Also from a supergravity perspective, the inclusion of massive modes is highly non-trivial; consistent truncations are hard to find, even when truncating to the massless Kaluza-Klein (KK) spectrum. In fact, until not long ago it was widely believed that consistency prevents one from keeping a finite number of massive KK modes. From the gauge/gravity correspondence perspective, in turn, the lower dimensional supergravity theories obtained from these reductions are assumed to possess field theory duals with various amounts of unbroken super(-conformal)symmetry.  Strikingly, the inclusion of charged operators on the field theory side, dual to massive bulk fields, opened the door for a stringy (``top-down") modelling of condensed matter phenomena, such as superfluidity and superconductivity and systems with non-relativistic conformal symmetries, via the holographic correspondence.  Even though the original work in these directions \cite{Gubser:2008px,Hartnoll:2008vx,Hartnoll:2008kx,Son:2008ye, Balasubramanian:2008dm} was based on a phenomenological, ``bottom-up" approach, it is clearly advantageous to consider top-down descriptions of these (or similar) systems. Indeed, a description in terms of ten or eleven-dimensional supergravity backgrounds may shed light on the existence of a consistent UV completion of the lower-dimensional effective bulk theories, while possibly fixing various parameters that appear to be arbitrary in the bottom-up constructions.

In this paper we shall be concerned with the consistent truncations of type IIB supergravity on squashed Sasaki-Einstein five-manifolds ($SE_{5}$) whose bosonic content was recently considered in \cite{Cassani:2010uw,Gauntlett:2010vu,Liu:2010sa}  (see \cite{Skenderis:2010vz} for related work). These constructions were largely motivated by the results of \cite{Maldacena:2008wh} ( see \cite{Herzog:2008wg, Adams:2008wt} also), which had a quite interesting by-product: while searching for solutions of type IIB supergravity with non-relativistic asymptotic symmetry groups, consistent five-dimensional truncations including massive bosonic modes were constructed. In particular, massive scalars arise from the breathing and squashing modes in the internal manifold, which is then a ``deformed" Sasaki-Einstein space, generalizing the case of breathing and squashing modes on spheres that had been studied in \cite{Bremer:1998zp, Liu:2000gk} (see also \cite{Buchel:2006gb}). Regarding the internal $SE_{5}$ manifold as a $U(1)$ bundle over a K\"ahler-Einstein ($KE$) base space of complex dimension two, the guiding principle behind these consistent truncations is to keep modes which are singlets only under the structure group of the $KE$ base. The bosonic sector of the corresponding truncations including massive modes in 11-$d$ supergravity on squashed $SE_{7}$ manifolds had been previously discussed in \cite{Gauntlett:2009zw}, and provided the basis for the embedding of the original holographic $AdS_{4}$ superconductors of \cite{Hartnoll:2008vx,Hartnoll:2008kx} into M-theory, a connection that was explored in \cite{Gauntlett:2009dn,Gauntlett:2009bh}. In our recent work \cite{Bah:2010yt} we have extended the consistent truncation of 11-$d$ supergravity on squashed $SE_{7}$ to include the fermionic sector, and in particular provided the effective 4-$d$ action describing the coupling of fermion modes to the M-theory holographic superconductor.

At the same time that the work of \cite{Gauntlett:2009dn} appeared, the embedding of an asymptotically $AdS_{5}$ holographic superconductor into type IIB supergravity was reported in \cite{Gubser:2009qm}. Continuing with the program we initiated in \cite{Bah:2010yt}, in the present work we discuss the extension of the consistent truncation of type IIB supergravity on $SE_{5}$ discussed in \cite{Cassani:2010uw,Gauntlett:2010vu,Liu:2010sa} to include the fermionic sector. In particular, as an application of our results we present the effective action describing the coupling of the fermion modes to the holographic superconductor of  \cite{Gubser:2009qm}. Knowing the precise form of said couplings is important from the point of view of the applications of gauge/gravity duality to the description of strongly coupled condensed matter phenomena, insofar as it determines the nature of fermionic correlators in the presence of superconducting condensates, that rely on how the fermionic operators of the dual theory couple to scalars. Hence, we set the stage for the discussion of these and related questions from a top-down perspective. A related problem involving a superfluid $p$-wave transition was studied in \cite{Ammon:2010pg}, in the context of (3+1)-dimensional supersymmetric field theories dual to probe $D5$-branes in $AdS^{5}\times S^5$. In the top-down approach starting from either ten or eleven-dimensional supergravity, inevitably the consistent truncations will include not only spin-1/2 fermions that might be of phenomenological interest but also spin-3/2 fields. One finds that these generally mix together via generalized Yukawa couplings, and this mixing will have implications for correlation functions in the dual field theory. One of our original motivations for the present work as well as  \cite{Bah:2010yt} was to understand this mixing in more detail and to investigate the existence of ``further truncations" which might involve (charged) spin-1/2 fermions alone. As we explain in section \ref{section:Examples}, in the present case we have indeed found such a model, containing a single spin-1/2 field, in the truncation corresponding to the type IIB holographic superconductor. 

This paper is organized as follows.  In section \ref{section:Ansatz}  we briefly review some aspects of the truncations of type IIB supergravity constructed in \cite{Cassani:2010uw,Gauntlett:2010vu,Liu:2010sa} and the extension of the bosonic ansatz to include the fermion modes. In section \ref{section:5d eqs} we present our main result: the effective five-dimensional action functional describing the dynamics of the fermions and their couplings to the bosonic fields. We chose to perform this calculation by directly reducing the 10-$d$ equations of motion for the gravitino and dilatino. The resulting action is consistent with 5-$d$ $N=4$ gauged supergravity, as has been anticipated.  In section \ref{section:N=4} we reduce the supersymmetry variation of the gravitino and dilatino, and comment on the supersymmetric structure of the five-dimensional theory by considering how the fermions fit into the supermultiplets of $N=4$ gauged supergravity. In principle, a complete mapping to the highly constrained form of $N=4$ actions could be made, although we do not give all of the details here. The $N=4$ theory has two vacuum $AdS_5$ solutions, one with $N=2$ supersymmetry and one without supersymmetry. In section \ref{section:vacua} we  linearize the fermionic sector in each of these vacua and demonstrate that as expected the gravitini attain masses via the St\"uckelberg mechanism, which is a useful check on the consistency of our results.  In section \ref{section:Examples} we apply our results to several further truncations of interest: the minimal gauged $N=2$ supergravity theory in five dimensions, and the dual \cite{Gubser:2009qm} of the $(3+1)$-dimensional holographic superconductor. We conclude in section \ref{section:Conclusions}. The details of many of our computations as well as a full accounting of our conventions appear in a series of appendices. 
\section{Type IIB supergravity on squashed Sasaki-Einstein five-manifolds}\label{section:Ansatz}
\subsection{Bosonic ansatz}
In this section we briefly review the ansatz for the bosonic fields in the consistent truncations of \cite{Cassani:2010uw,Gauntlett:2010vu,Liu:2010sa}. In the following subsection, we will discuss the extension of this ansatz to include the fermionic fields of type IIB supergravity. 
Here we mostly follow the type IIB conventions of \cite{Gauntlett:2010vu,Gauntlett:2005ww,Gauntlett:2007ma},  with slight modifications as we find appropriate. Further details of these conventions can be found in appendix \ref{Appendix:Conventions}. 

The Kaluza-Klein metric ansatz in the truncations of interest is given by \cite{Cassani:2010uw,Gauntlett:2010vu,Liu:2010sa}
\begin{equation}\label{metric ansatz}
ds^2_{10}=e^{2W(x)}ds_{E}^2(M)+e^{2U(x)}ds^2(KE)+e^{2V(x)}\bigl(\eta +A_{1}(x)\bigr)^2\, ,
\end{equation}

\noindent where $W(x) = -\frac{1}{3}(4U(x) + V(x))$. Here, $M$ is an arbitrary ``external" five-dimensional manifold, with coordinates denoted generically by $x$ and five-dimensional Einstein-frame metric $ds_{E}^{2}(M)$, and $KE$ is an ``internal" four-dimensional K\"ahler-Einstein manifold (henceforth referred to as ``KE base") coordinatized by $y$ and possessing K\"ahler form $J$. The one-form $A_{1}$ is defined in $T^*M$ and $\eta \equiv d\chi + {\cal A}(y)$, where ${\cal A}$ is an element of $T^*KE$ satisfying $d{\cal A} \equiv {\cal F} = 2J$. For a fixed point in the external manifold, the compact coordinate $\chi$ parameterizes the fiber of a $U(1)$ bundle over $KE$, and the five-dimensional internal manifold spanned by $(y,\chi)$ is then a squashed Sasaki-Einstein manifold, with the breathing and squashing modes parameterized by the scalars $U(x)$ and $V(x)$.\footnote{In particular, $U-V$ is the squashing mode, describing the squashing of the $U(1)$ fiber with respect to the KE base, while the breathing mode $4U + V$ modifies the overall volume of the internal manifold. When $U=V=0$, the internal manifold becomes a five-dimensional Sasaki-Einstein manifold $SE_{5}$.} In addition to the metric, the bosonic content of type IIB supergravity \cite{Schwarz:1983qr,Howe:1983sra} includes the dilaton $\Phi$, the NSNS 3-form field strength $H_{(3)}$, and the RR field strengths $F_{(1)} \equiv dC_{0}$, $F_{(3)}$ and $F_{(5)}$, where $C_{0}$ is the axion and $F_{(5)}$ is self-dual. The rationale behind the corresponding ans\"atze is the idea that the consistency of the dimensional reduction is a result of truncating the KK tower to include fields that transform as singlets only under the structure group of the KE base, which in this case corresponds to $SU(2)$. This prescription allows for an interesting spectrum in the lower dimensional theory, inasmuch as the $SU(2)$ singlets include fields that are charged under the $U(1)$ isometry generated by $\partial_{\chi}$. The globally defined K\"ahler 2-form $J = d{\cal A}/2$ and the holomorphic $(2,0)$-form $\Sigma_{(2,0)}$ define the K\"ahler and complex structures, respectively, on the KE base. They are $SU(2)$-invariant and can be used in the reduction of the various fields to five dimensions. The $U(1)$-bundle over $KE$ is such that they satisfy
\begin{equation}
\Sigma_{(2,0)} \wedge \Sigma_{(2,0)}^{*} = 2J^{2}\, ,\qquad \mbox{and}\qquad d\Sigma_{(2,0)} = 3i {\cal A}\wedge \Sigma_{(2,0)}\, .
\end{equation}
More precisely, as will be clear from the discussion to follow below, the relevant charged form $\Omega$ on the total space of the bundle that should enter the ansatz for the various form fields is given by
\begin{equation}
\Omega \equiv e^{3i\chi}\Sigma_{(2,0)}\, ,
\end{equation}
and satisfies
\begin{equation}
d\Omega = 3i\eta \wedge\Omega \, .
\end{equation}

\noindent The ans\"atze for the bosonic fields is then \cite{Gauntlett:2010vu}
\begin{eqnarray} \label{KKformshk}
F_{(5)} &=&
4 e^{8W+Z} \textrm{vol}^E_5 
+e^{4(W+U)} * K_2 \wedge J+ K_1 \wedge J \wedge J\nonumber\\
&&+ \left[
2e^Z J \wedge J
-2e^{-8U} *K_1   +K_2 \wedge J \right]\wedge (\eta+A_1)\nonumber\\
 && + \left[ e^{4(W+U)} *L_2 \wedge \Omega + L_2 \wedge \Omega \wedge (\eta+A_1) +\mbox{c.c.} \right]
\label{defF5red} \\[8pt]
F_{(3)} &=& G_3 +G_2 \wedge (\eta+A_1) +G_1 \wedge J +G_0\, J\wedge (\eta+A_1)
\nonumber\\
&&
+\Bigl[ N_1 \wedge \Omega  +N_0\,\Omega\wedge (\eta+A_1)+\mbox{c.c.}\Bigr]
\\[8pt]
 H_{(3)} &=& H_3 +H_2 \wedge (\eta+A_1) +H_1 \wedge J +H_0\,  J\wedge (\eta+A_1)
\nonumber\\
&&
+\Bigl[ M_1 \wedge \Omega  +M_0\,\Omega\wedge (\eta+A_1)+\mbox{c.c.}\Bigr]
 \\[8pt]
C_{(0)} &=& a
\\[8pt]
\Phi &=& \phi
\end{eqnarray}

\noindent  where $\textrm{vol}^E_5$ and $*$ are the volume form and Hodge dual appropriate to the five-dimensional Einstein-frame metric $ds^{2}_{E}(M)$, and $W(x) = -\frac{1}{3}(4U(x) + V(x))$ as before. Several comments are in order. First, all the fields other than $(\eta, J, \Omega)$ are defined on $\sfLam^* T^{*}M$. $Z$, $a$, $\phi$, $G_{0}$, $H_{0}$ are real scalars, and $M_{0}$, $N_{0}$ are complex scalars. The form fields $G_{1}$, $G_{2}$, $G_{3}$, $H_{1}$, $H_{2}$, $H_{3}$, $K_{1}$ and $K_{2}$ are real, while $M_{1}$, $N_{1}$ and $L_{2}$ are complex forms. As pointed out in \cite{Gauntlett:2010vu}, the scalars $G_{0}$ and $H_{0}$ vanish by virtue of the type IIB Bianchi identities. We also notice that the self-duality of $F_{(5)}$ is automatic in the ansatz \eqref{defF5red}: the first two lines are duals of each other, while the last line is self-dual. 

Inserting the ansatz into the type IIB equations of motion and Bianchi identities (Appendix \ref{Appendix:typeIIB}), one finds that the various fields are related as\footnote{We have chosen the notation of Ref. \cite{Gauntlett:2010vu} apart from replacing their $\chi,\xi$ with $X,Y$, to avoid confusion with the fiber coordinate.}
\beqn \label{fieldstr1}
H_3&=&dB_2 +\frac12(db-2B_1) \wedge F_2 \nonumber\\
G_3&=& dC_2 -adB_2 +\frac12\left(dc-adb-2C_1+2 a B_1\right) \wedge  F_2\nonumber\\
H_2&=&dB_1\nonumber\\
F_2&=&dA_1\nonumber\\
G_2&=&dC_1-adB_1\nonumber\\
K_2 &=&  dE_1+\frac12(db-2B_1)\wedge (dc-2C_1)  \nonumber\\
G_1&=& dc-adb-2C_1+2aB_1 \nonumber\\
H_1&=&db-2B_1\nonumber\\
K_1&=&dh-2E_1-2A_1+ Y^*DX+ Y DX^*-X D Y^*-X^*D Y\nonumber\\
M_1&=&D Y\nonumber\\
N_1&=&DX-aD Y\nonumber\\
M_0&=&3i Y\nonumber\\
N_0&=&3i(X-a Y)\nonumber\\
e^Z&=&1+3i( Y^*X- Y X^*),
\end{eqnarray}

\noindent where $F_{2}\equiv dA_{1}$, $X, Y$ and $L_2,M_1,N_1$ are complex, and $D Y=d Y-3iA_1 Y$, $DX=dX-3iA_1X$.

As was explained in detail in  \cite{Cassani:2010uw,Gauntlett:2010vu}, the physical scalars parameterize the coset $SO(1,1)\times \bigl( SO(5,2)/(SO(5)\times SO(2))\bigr)$, while the structure of the 1-forms and 2-forms is such that a $Heis_3\times U(1)$ subgroup is gauged.

\subsection{Fermionic ansatz}
The fermionic content of type IIB supergravity comprises a positive chirality dilatino and a negative chirality gravitino. Instead of expressing the theory in terms of pairs of Majorana-Weyl fermions, we find it notationally simplest to use complex Weyl spinors. Quite generally, we would like to decompose the gravitino using an ansatz of the form
\begin{align}
\Psi_a(x,y,\chi) = {}&\sum_I\psi_a^I(x)\otimes \eta^I(y,\chi)\\
\Psi_\alpha(x,y,\chi)={}&\sum_I\lambda^I(x)\otimes \eta_\alpha^I(y,\chi)\\
\Psi_\f(x,y,\chi)={}&\sum_I\varphi^I(x)\otimes \eta_{\f}^I(y,\chi)\, ,
\end{align}

\noindent where $a$, $\alpha$ and $\f$ denote the indices in the direction of the external manifold, the KE base, and the fiber, respectively. The projection to singlets under the structure group of the KE base was recently described in great detail for the case of $D=11$ supergravity compactified on squashed $SE_{7}$ manifolds \cite{Bah:2010yt}. Since the principles at work in the present case are essentially the same, here we limit ourselves to pointing to a few relevant facts and results. As we have discussed, the five-dimensional internal space is the total space of a $U(1)$ bundle over a KE base. In general, the base is not spin, and therefore spinors do not necessarily exist globally on the base. However, it is always possible to define a $Spin^{c}$ bundle globally on $KE$ (see  \cite{Martelli:2006yb}, for example), and our (c-)spinors will then be sections of this bundle. Indeed, we have seen above that the holomorphic form $\Omega$ is also charged under this $U(1)$. The $U(1)$ generator is proportional to $\pa_\chi$, and hence $\nabla_\alpha- {\cal A}_\alpha\pa_\chi$ is the gauge connection on the $Spin^c$ bundle, where $\nabla_{\alpha}$ is the covariant derivative on $KE$. Of central importance to us in the reduction to invariants of the structure group are the gauge-covariantly-constant spinors, which can be defined on any K\"ahler manifold \cite{Hitchin19741} and satisfy in the present context
\beq\label{gauge cov const spinor eq}
(\nabla_\alpha- {\cal A}_\alpha\pa_\chi)\varepsilon(y,\chi)=0\, ,
\eeq
where
\begin{equation}\label{gauge cov const spinor}
\varepsilon(y,\chi)=\varepsilon(y) e^{ie\chi}
\end{equation}
for fixed ``charge" $e$. For a KE base of real dimension $d_{b}$, these satisfy (see  \cite{Bah:2010yt},\cite{Gibbons:2002th} for example)\footnote{All of our Clifford algebra and spinor conventions are compiled in Appendix \ref{Appendix:Conventions}.}
\beq
Q\varepsilon \equiv -iJ_{\alpha\beta}\Gamma^{\alpha\beta}\varepsilon = \frac{4ed_{b}}{d_{b}+2} \varepsilon\, .
\eeq

\noindent In other words, the matrix $Q=-iJ_{\alpha\beta}\Gamma^{\alpha\beta}$ on the left is (up to normalization) the U(1) charge operator. It has maximum eigenvalues $\pm d_{b}$, and the corresponding spinors have charge 
\beq
e=\pm \frac{d_{b} + 2}{4}\, .
\eeq

\noindent These two spinors are charge conjugates of one another, and we will henceforth denote them by $\varepsilon_{\pm}$. By definition, they satisfy $\cFsl\varepsilon_{\pm} = iQ\varepsilon_{\pm} = \pm id_{b}\,\varepsilon_{\pm}$, where $\cFsl \equiv (1/2){\cal F}_{\alpha\beta}\Gamma^{\alpha\beta}$. These spinors with maximal $Q$-charge are in fact the singlets under the structure group, and they constitute the basic building blocks of the reduction ansatz for the fermions. In the case at hand $d_{b}=4$ and the structure group is $SU(2)$; in fact we have an unbroken $SU(2)_L\times U(1)$ subgroup of $Spin(4)$ in which the spinor transforms as  ${\bf 2}_0\oplus {\bf 1}_+\oplus {\bf 1}_-$. In the complex basis introduced in \ref{Appendix:Conventions-Fluxes}, we find
\beq
Q_\alpha\varepsilon_\pm=\pm\frac12\varepsilon_\pm \quad (\alpha=1,2)
\eeq
and
\beq
\bar P_\alpha\varepsilon_+=0,\ \ \ \ P_\alpha\varepsilon_-=0\, ,
\eeq

\noindent where $Q_\alpha=\Gamma^{\alpha\bar\alpha}$, $P_\alpha=\Gamma^\alpha\Gamma^{\bar\alpha}$, and $\bar P_\alpha=\Gamma^{\bar\alpha}\Gamma^\alpha$. In the Fock state basis, these are $\varepsilon_\pm\leftrightarrow |\pm\frac12,\pm\frac12\rangle$ and the remaining two states form a (charge-zero) doublet. Unlike the two $SU(3)$ singlet spinors that were used to reduce the gravitino in the 11-$d$ case, here the two singlets have the {\it same} chirality in $4+0$ dimensions, that is $\gamma_\f \varepsilon_\pm=\varepsilon_\pm$ (this follows, since $\gamma_\f =-\gamma^{1234}=\prod_{\alpha}2Q_\alpha$). Similarly, for the complex form $\Sigma_{(2,0)}$ we find $[Q,\slashed{\Sigma}]=8\slashed{\Sigma}$, which means that $\Sigma_{(2,0)}$ carries charge $e_{\Sigma} = 3$ and justifies the definition $\Omega = e^{3i\chi}\Sigma_{(2,0)}$ discussed above.

 We are now in position to write the reduction ansatz for the gravitino and dilatino. Dropping all the $SU(2)$ representations other than the singlets, we take
\beqn
\label{fermion ansatz first}\Psi_a(x,y,\chi)&=&\psi^{(+)}_a(x)\otimes \varepsilon_+(y)e^{\frac32 i\chi}\otimes u_-+\psi^{(-)}_a(x)\otimes \varepsilon_-(y)e^{-\frac32 i\chi}\otimes u_-\\
\Psi_\alpha(x,y,\chi)&=&\rho^{(+)}(x)\otimes  \gamma_\alpha\varepsilon_+(y)e^{\frac32 i\chi}\otimes u_-\\
\Psi_{\bar\alpha}(x,y,\chi)&=&\rho^{(-)}(x)\otimes  \gamma_{\bar\alpha}\varepsilon_-(y)e^{-\frac32 i\chi}\otimes u_-\\
\Psi_\f(x,y,\chi)&=&\varphi^{(+)}(x)\otimes \varepsilon_+(y)e^{\frac32 i\chi}\otimes u_-+\varphi^{(-)}(x)\otimes \varepsilon_-(y)e^{-\frac32 i\chi}\otimes u_-\\
\lambda(x,y,\chi)&=&\lambda^{(+)}(x)\otimes \varepsilon_+(y)e^{\frac32 i\chi}\otimes u_++\lambda^{(-)}(x)\otimes \varepsilon_-(y)e^{-\frac32 i\chi}\otimes u_+ \label{fermion ansatz last}
\eeqn

\noindent where $\varphi^{(\pm)}, \rho^{(\pm)}$ and $\psi_a^{(\pm)}$ are $(4+1)$-dimensional spinors on $M$, the superscript $\conj$ denotes charge conjugation, and we have used the complex basis introduced in \ref{Appendix:Conventions-Fluxes} for the KE base directions ($\alpha, \bar{\alpha} = 1,2$).  The constant spinors $ u_+=\binom{1}{0}$ and $ u_-=\binom{0}{1}$ have been introduced as bookkeeping devices to keep track of the $D=10$ chiralities. Since our starting spinors were only Weyl in $D=10$ (as opposed to Majorana-Weyl) there is no relation between, say, $\lambda^{(+)}$ and $\lambda^{(-)}$; they are independent Dirac spinors in $4+1$ dimensions, and the same applies to the rest of the spinors in the ansatz.  Although one could write the $(4+1)$-spinors as symplectic Majorana, there is no real benefit to introducing such notation at this point in the discussion. Notice that all of these modes are annihilated by the gauge-covariant derivative on $KE$. Equations \eqref{fermion ansatz first}-\eqref{fermion ansatz last} provide the starting point for the dimensional reduction of the $D=10$ equations of motion of type IIB supergravity down to $d=5$.

According to the charge conjugation conventions in \ref{Appendix:ChargeConjugation}, we also find
\beqn
\Psi_a^\conj(x,y,\chi)&=&
\psi^{(-)\conj}_a(x)\otimes \varepsilon_+(y)e^{\frac32 i\chi}\otimes u_-
-\psi^{(+)\conj}_a(x)\otimes \varepsilon_-(y)e^{-\frac32 i\chi}\otimes u_-
\\
(\Psi_\alpha)^\conj(x,y,\chi)&=&-\rho^{(+)\conj}(x)\otimes  \gamma_{\bar\alpha}\varepsilon_-(y)e^{-\frac32 i\chi}\otimes u_-
\\
(\Psi_{\bar\alpha})^\conj(x,y,\chi)&=&\rho^{(-)\conj}(x)\otimes  \gamma_{\alpha}\varepsilon_+(y)e^{\frac32 i\chi}\otimes u_-
\\
\Psi_\f^\conj(x,y,\chi)&=&
\varphi^{(-)\conj}(x)\otimes \varepsilon_+(y)e^{\frac32 i\chi}\otimes u_-
-\varphi^{(+)\conj}(x)\otimes \varepsilon_-(y)e^{-\frac32 i\chi}\otimes u_-
\\
\lambda^\conj(x,y,\chi)&=&
-\lambda^{(-)\conj}(x)\otimes \varepsilon_+(y)e^{\frac32 i\chi}\otimes u_+
+\lambda^{(+)\conj}(x)\otimes \varepsilon_-(y)e^{-\frac32 i\chi}\otimes u_+
\eeqn

\section{Five-dimensional equations of motion and effective action}\label{section:5d eqs}

The type IIB fermionic equations of motion to linear order in the fermions are given by (see appendix \ref{Appendix:typeIIB} for details)
\beqn
\hat\cDsl\lambda &=& \frac{i}{8}\Fsl_{(5)}\lambda + {\cal O}(\Psi^{2})\\
\Gamma^{ABC}\hat {\cal D}_B\Psi_C
&=&-\frac{1}{8}\Gsl^*\Gamma^A\lambda+\fudge \frac{1}{2}\Psl\Gamma^A\lambda^\conj + {\cal O}(\Psi^{3})
\eeqn

\noindent Here, $\hat{D}$ denotes the flux-dependent supercovariant derivative, which acts as follows:
\begin{eqnarray}
\label{DilSuper}
\hat{\cDsl}\lambda &=& \left(\hat{\slashed{\nabla}} -\frac{3i}{2}\slashed{Q}\right)\lambda -\frac{1}{4}\Gamma^A\Gsl\Psi_A-\Gamma^A\Psl\Psi^\conj_A \,, \\ 
\hat {\cal D}_B\Psi_C &=& \left(\hat{\nabla}_B -\frac{i}{2}Q_B\right)\Psi_C+\frac{i}{16}\Fsl_{(5)}\Gamma_B\Psi_C-\frac{1}{16}S_{B}\Psi^\conj_C \, ,
\end{eqnarray}

\noindent where $\hat{\nabla}_{B}$ denotes the ordinary 10-$d$ covariant derivative and we have defined 
\begin{eqnarray}
\label{definition S}
S_B \equiv\frac{1}{6}\left({\Gamma_B}^{DEF}G_{DEF}-9\Gamma^{DE}G_{BDE}\right).
\end{eqnarray}

\noindent As described in Appendix \ref{Appendix:typeIIBbosonic}, defining the axion-dilaton $\tau = C_{(0)} + ie^{-\Phi}=a + ie^{-\phi}$ our conventions imply
\begin{equation}\label{definition G}
G = ie^{\Phi/2}\left(\tau dB - dC_{(2)}\right) =  -\left(e^{-\phi/2}H_{(3)} + ie^{\phi/2}F_{(3)}\right),
\end{equation}

\noindent and 
\begin{equation}
P = \frac{i}{2}e^{\Phi}d\tau = \frac{d\phi}{2} + \frac{i}{2}e^{\phi}da\, ,\qquad Q = -\frac{1}{2}e^{\Phi}dC_{(0)}=-\frac{1}{2}e^{\phi}da\, .
\end{equation}

\noindent It will prove convenient to introduce a compact notation as follows:
\begin{align}
\mathcal{G}_{1} &= e^{\frac{1}{2}\left(\phi - 4U\right)}\left(G_{1} - ie^{-\phi}H_{1}\right)
&
\tilde{\mathcal{G}}_{1} &= e^{\frac{1}{2}\left(\phi - 4U\right)}\left(G_{1} +ie^{-\phi}H_{1}\right)
\\
\mathcal{G}_{2} &= e^{\frac{1}{2}\left(\phi + 4U\right)}\Sigma\left(G_{2} - ie^{-\phi}H_{2}\right)
&
\tilde{\mathcal{G}}_{2} &= e^{\frac{1}{2}\left(\phi + 4U\right)}\Sigma\left(G_{2} +ie^{-\phi}H_{2}\right)
\\
\mathcal{G}_{3} &= e^{\frac{1}{2}\left(\phi + 4U\right)}\Sigma^{-1}\left(G_{3} - ie^{-\phi}H_{3}\right)
&
\tilde{\mathcal{G}}_{3} &= e^{\frac{1}{2}\left(\phi + 4U\right)}\Sigma^{-1}\left(G_{3} +ie^{-\phi}H_{3}\right)
\\
\mathcal{N}_{1}^{(+)} &= e^{\frac{1}{2}\left(\phi - 4U\right)}\left(N_{1} - ie^{-\phi}M_{1}\right)
&
\tilde{\mathcal{N}}_{1}^{(+)} &= e^{\frac{1}{2}\left(\phi - 4U\right)}\left(N_{1} +ie^{-\phi}M_{1}\right)
\\
\mathcal{N}_{1}^{(-)} &= e^{\frac{1}{2}\left(\phi - 4U\right)}\left(N^*_{1} - ie^{-\phi}M^*_{1}\right)
&
\tilde{\mathcal{N}}_{1}^{(-)} &= e^{\frac{1}{2}\left(\phi - 4U\right)}\left(N^*_{1} +ie^{-\phi}M^*_{1}\right)
\\
\mathcal{N}_{0}^{(+)} &= e^{\frac{1}{2}\left(\phi - 4U\right)}\Sigma^{2}\left(N_{0} - ie^{-\phi}M_{0}\right)
&
\tilde{\mathcal{N}}_{0}^{(+)} &= e^{\frac{1}{2}\left(\phi - 4U\right)}\Sigma^{2}\left(N_{0} +ie^{-\phi}M_{0}\right)
\\
\mathcal{N}_{0}^{(-)} &= e^{\frac{1}{2}\left(\phi - 4U\right)}\Sigma^{2}\left(N^*_{0} - ie^{-\phi}M^*_{0}\right)
&
\tilde{\mathcal{N}}_{0}^{(-)} &= e^{\frac{1}{2}\left(\phi - 4U\right)}\Sigma^{2}\left(N^*_{0} +ie^{-\phi}M^*_{0}\right)
\end{align}

\noindent where the scalar $\Sigma$ is defined as
$\Sigma \equiv  e^{2(W + U)} = e^{-\frac{2}{3}(U + V)}$. Its significance will be reviewed later in the paper.

The detailed derivation of the equations of motion is performed in Appendix \ref{Appendix:EOM}, and we will not reproduce them here in the main body of the paper as the expressions are lengthy. Given those equations of motion, we will write an action from which they may be derived. Before doing so, we first consider the kinetic terms and introduce a field redefinition such that the kinetic terms are diagonalized. 

\subsection{Field redefinitions}
In order to find the appropriate field redefinitions it is enough to consider the derivative terms, which follow from a Lagrangian density of the form (with respect to the 5-$d$ Einstein frame-measure $d^{5}x\sqrt{-g_{5}^{E}}\,$ )
\begin{align}
L_{kin}^{(\pm)} =
{}&
e^{W}\biggl[\frac{1}{2}\bar{\lambda}^{(\pm)}\Dsl \lambda^{(\pm)}+\bar{\psi}^{(\pm)}_{a}\left(\gamma^{abc}D_{b}\psi_{c}^{(\pm)}-4i\gamma^{ab}D_{b}\rho^{(\pm)}-i\gamma^{ab}D_{b}\varphi^{(\pm)}\right)
\nonumber\\
&\hphantom{e^{W}\Bigl[}
 -i\bar{\rho}^{(\pm)}\left(4 \gamma^{ab}D_{a}\psi_{b}^{(\pm)}-12i\Dsl\,\rho^{(+)}-4i \Dsl\,\varphi^{(\pm)}\right)
\nonumber\\
&\hphantom{e^{W}\Bigl[}
+ \bar{\varphi}^{(\pm)}\left(-i\gamma^{ab}D_{a}\psi^{(\pm)}_{b}-4\Dsl\, \rho^{(\pm)} \right)\biggr].
\end{align} 

\noindent Shifting the gravitino as\footnote{To avoid confusion, we note that the notation $\bar\varphi^{(\pm)}$ means $(\varphi^{(\pm)})^\dagger\gamma^0$, etc.}
\begin{equation}
\psi^{(\pm)}_{a} = \tilde{\psi}^{(\pm)}_{a} + \frac{i}{3}\gamma_{a}\left(\varphi^{(\pm)} + 4\rho^{(\pm)}\right)
 \quad \Rightarrow \quad 
 \bar{\psi}^{(\pm)}_{a} = \bar{\tilde{\psi}}^{(\pm)}_{a} + \frac{i}{3}\left(\bar{\varphi}^{(\pm)} + 4\bar{\rho}^{(\pm)}\right) \gamma_{a}\, ,
\end{equation}

\noindent we obtain
\begin{align}
L_{kin}^{(\pm)} ={}& e^{W}\biggl[\frac{1}{2}\bar{\lambda}^{(\pm)}\Dsl \lambda^{(\pm)}+\bar{\tilde{\psi}}^{(\pm)}_{a}\gamma^{abc}D_{b}\tilde{\psi}_{c}^{(\pm)} +8\bar{\rho}^{(\pm)}\Dsl\, \rho^{(\pm)}
\nonumber\\
&\hphantom{e^{W}\biggl[}
+ \frac{4}{3}\left(\bar{\rho}^{(\pm)} + \bar{\varphi}^{(\pm)}\right)\Dsl\, \left(\rho^{(\pm)}+\varphi^{(\pm)}\right)\biggr].
\end{align}

\noindent Then we are led to define\footnote{One should not confuse the one-form $\eta$ dual to the Reeb vector field with the fermions $\eta^{(\pm)}$. }
\begin{align}
\tilde{\lambda}^{(\pm)} &= e^{W/2}\lambda^{(\pm)}\label{canonical fields 1}
\\
\zeta^{(\pm)}_{a} &= e^{W/2}\left[\psi^{(\pm)}_{a}-\frac{i}{3}\gamma_{a}\left(\varphi^{(\pm)} + 4\rho^{(\pm)}\right)\right]\label{canonical fields 2}
\\
\xi^{(\pm)} &= 4e^{W/2}\rho^{(\pm)}\label{canonical fields 3}
\\
\eta^{(\pm)} &= 2e^{W/2}\left(\rho^{(\pm)}+\varphi^{(\pm)}\right),\label{canonical fields 4}
\end{align}

\noindent which results in
\begin{align}
L_{kin}^{(\pm)} = {}&\frac{1}{2}\bar{\tilde{\lambda}}^{(\pm)}\Dsl \tilde{\lambda}^{(\pm)} + \bar{\zeta}^{(\pm)}_{a}\gamma^{abc}D_{b}\zeta_{c}^{(\pm)} +\frac{1}{2}\bar{\xi}^{(\pm)}\Dsl\, \xi^{(\pm)} 
+ \frac{1}{3}\bar{\eta}^{(\pm)}\Dsl\, \eta^{(\pm)}\label{fermionnormsone}
\\
&
- \frac{1}{2}\left[\bar{\zeta}^{(\pm)}_{a}\gamma^{abc}\left(\partial_{b}W\right)\zeta_{c}^{(\pm)}
 +\frac{1}{2}\bar{\xi}^{(\pm)}\left(\pasl W\right) \xi^{(\pm)} + \frac{1}{3}\bar{\eta}^{(\pm)}\left(\pasl W\right) \eta^{(\pm)} \right].\label{fermionnormstwo}
\end{align}

\noindent The $W$-dependent interaction terms in the second line are produced by the action of the derivatives on the warping factors involved in the field redefinitions, and they will cancel against similar terms in the interaction Lagrangian. We note that the fields we have defined are not canonically normalized. We have done this simply to avoid square-root factors.

The equations of motion written in terms of the fields \eqref{canonical fields 1}-\eqref{canonical fields 4} are given explicitly in Appendix \ref{Appendix:EOM}. They follow from an effective $d=5$ action that we derive below.

\subsection{Effective action}
The equations of motion for the 5d fields \eqref{canonical fields 1}-\eqref{canonical fields 4}, which are explicitly displayed in appendix \ref{Appendix:EOM}, follow from an effective action functional of the form
\begin{align}\label{effective 5d action}
S_{4+1} = K_{5}\int d^{5}x\sqrt{-g_{5}^E}\biggl[&\frac{1}{2}\bar{\tilde{\lambda}}^{(+)}\Dsl \tilde{\lambda}^{(+)} + \bar{\zeta}^{(+)}_{a}\gamma^{abc}D_{b}\zeta_{c}^{(+)} +\frac12\bar{\xi}^{(+)}\Dsl\, \xi^{(+)} 
+ \frac{1}{3}\bar{\eta}^{(+)}\Dsl\, \eta^{(+)}
\nonumber\\
&
+ \frac{1}{2}\bar{\tilde{\lambda}}^{(-)}\Dsl \tilde{\lambda}^{(-)} + \bar{\zeta}^{(-)}_{a}\gamma^{abc}D_{b}\zeta_{c}^{(-)} +\frac12\bar{\xi}^{(-)}\Dsl\, \xi^{(-)} 
+ \frac{1}{3}\bar{\eta}^{(-)}\Dsl\, \eta^{(-)}
\nonumber\\
&
+ \mathcal{L}_{\bar{\psi}\psi}^{(+)}+\mathcal{L}_{\bar{\psi}\psi}^{(-)} +\frac{1}{2}\left(\mathcal{L}_{\bar{\psi}\psi^\conj}^{(+)}+\mathcal{L}_{\bar{\psi}\psi^\conj}^{(-)} + \mbox{ c.c.}\right)\biggr]
\end{align}

\noindent where $K_{5}$ is a normalization constant depending on the volume of the $KE$ base, the length of the fiber parameterized by $\chi$, and the normalization of the spinors $\varepsilon_{\pm}$.  Here, $D_{a} \psi^{(\pm)}=\left(\nabla_{a} \mp\frac{3i}{2}A_{1a}\right)\psi^{(\pm)}$ for $\psi = \tilde{\lambda}, \psi_{a},\eta,\xi$, and the interaction Lagrangians are given by
\begin{equation}
\mathcal{L}_{\bar{\psi}\psi}^{(\pm)} = \mathcal{L}_{mass}^{(\pm)} +\mathcal{L}_{1}^{(\pm)} + \mathcal{L}_{2}^{(\pm)}
\end{equation}

\noindent where we have defined
\begin{align}
\mathcal{L}_{mass}^{(\pm)}
 ={}&
 \mp \frac{1}{2}\left(e^{-4U}\Sigma^{-1}+\frac32\Sigma^2\pm e^{Z+4W}\right)\bar{\tilde{\lambda}}^{(\pm)}\tilde{\lambda}^{(\pm)}
\mp\left(e^{-4U}\Sigma^{-1}+\frac32\Sigma^2\mp e^{Z+4W}\right)\bar{\zeta}_{a}^{(\pm)}\gamma^{ac}\zeta^{(\pm)}_{c}
\nonumber\\
&
\mp\frac{1}{9}\left(e^{-4U}\Sigma^{-1}- \frac{15}{2}\Sigma^2\pm5e^{Z+4W}\right)\bar{\eta}^{(\pm)}\eta^{(\pm)}
\pm \frac32\left(e^{-4U}\Sigma^{-1}-\frac12\Sigma^{2}\mp e^{Z+4W}\right)\bar{\xi}^{(\pm)}\xi^{(\pm)}
\nonumber\\
&
\pm\frac{1}{3}i\left(e^{-4U}\Sigma^{-1}-3\Sigma^2\pm2e^{Z+4W}\right)
\left(\bar{\zeta}_{a}^{(\pm)}\gamma^{a}\eta^{(\pm)}+\bar{\eta}^{(\pm)}\gamma^{a}\zeta^{(\pm)}_{a}\right)
\nonumber\\
&
\mp \frac{ 2}{3}\left(e^{-4U}\Sigma^{-1}\pm2e^{Z+4W}\right)
\left(\bar{\eta}^{(\pm)}\xi^{(\pm)}+\bar{\xi}^{(\pm)}\eta^{(\pm)}\right)
\nonumber\\
&
\mp i\left(  e^{-4U}\Sigma^{-1}\mp e^{Z+4W}\right)
\left(\bar{\zeta}_{a}^{(\pm)}\gamma^{a}\xi^{(\pm)}+\bar{\xi}^{(\pm)}\gamma^a\zeta^{(\pm)}_{a} \right)
\nonumber\\
&
\pm\cnpm{0}{}{}\left[
\frac12\bar{\tilde{\lambda}}^{(\pm)}\gamma^a\zeta_a^{(\mp)}
+\frac23 i\bar{\tilde{\lambda}}^{(\pm)}\eta^{(\mp)}
+\frac12i\bar{\tilde{\lambda}}^{(\pm)}\xi^{(\mp)}
\right]
\nonumber\\
&
 \pm\cntpm{0}{}{}\left[
 \frac12\bar{\zeta}_{a}^{(\pm)}\gamma^a\tilde{\lambda}^{(\mp)}
  +\frac{2}{3}i \bar{\eta}^{(\pm)}\tilde{\lambda}^{(\mp)}
+\frac12i\bar{\xi}^{(\pm)}\tilde{\lambda}^{(\mp)}
 \right]
\end{align}
\begin{align}
\mathcal{L}_{1}^{(\pm)}
 ={}&
 +\frac{1}{8}i \bar{\tilde{\lambda}}^{(\pm)} \left[3e^{\phi}(\pasl a)+2e^{-4U} \slashed{K}_1\right]\tilde{\lambda}^{(\pm)}
+ \frac14i e^{-4U} \bar{\zeta}_{a}^{(\pm)}\left(
e^{\phi} \gamma^{abc}(\pa_b a)+2\gamma^{[c}\slashed K_1\gamma^{a]}
 \right)\zeta^{(\pm)}_{c}
 \nonumber\\
&
 +\frac18i\bar{\xi}^{(\pm)} \Bigl[e^{\phi}(\pasl a)+6e^{-4U} \slashed K_1\Bigr]\xi^{(\pm)}
+\frac{1}{12}i\bar{\eta}^{(\pm)} \Bigl[e^{\phi}(\pasl a)-2e^{-4U}\slashed{K}_1\Bigr]\eta^{(\pm)} 
\nonumber
\\
& 
+\bar{\zeta}_{a}^{(\pm)}\Bigl(i(\pasl U)-\frac12e^{-4U} \slashed K_1\Bigr)\gamma^{a}\xi^{(\pm)}
+\bar{\xi}^{(\pm)}\gamma^{a}\Bigl(-i(\pasl U)- \frac12e^{-4U}\slashed{K}_1\Bigr)\zeta^{(\pm)}_{a} 
\nonumber\\
&
-\frac12i\bar{\zeta}_{a}^{(\pm)}(\Sigma^{-1}\pasl\Sigma)\gamma^{a}\eta^{(\pm)}
+\frac{1}{2}i\bar{\eta}^{(\pm)}\gamma^{a}(\Sigma^{-1}\pasl\Sigma)\zeta^{(\pm)}_{a}
\nonumber\\
&
\pm\frac{1}{2}i\bar{\tilde{\lambda}}^{(\pm)}\gamma^a \cnpmsl{1}\zeta^{(\mp)}_{a}  
 \pm\frac12i\bar{\zeta}_{a}^{(\pm)}\cntpmsl{1}\gamma^a\tilde{\lambda}^{(\mp)}
\pm\frac12\bar{\tilde{\lambda}}^{(\pm)}\cnpmsl{1}\xi^{(\mp)} 
\pm\frac12\bar{\xi}^{(\pm)}\cntpmsl{1}\tilde{\lambda}^{(\mp)}
\nonumber\\
& 
\pm \frac{1}{4}i\left(\bar{\tilde{\lambda}}^{(\pm)}\cgsl{1}\xi^{(\pm)}+\bar{\xi}^{(\pm)}\cgtsl{1}\tilde{\lambda}^{(\pm)} \right)
\mp\frac{1}{4}\left(\bar{\tilde{\lambda}}^{(\pm)}\gamma^a\cgsl{1}\zeta^{(\pm)}_{a}+\bar{\zeta}_{a}^{(\pm)}\cgtsl{1}\gamma^a\tilde{\lambda}^{(\pm)}\right)
\end{align}

\noindent and
\begin{align}
\mathcal{L}_{2}^{(\pm)}
 ={}&
+\frac{1}{8}\bar{\tilde{\lambda}}^{(\pm)}\gamma^a\left(i\cgsl{3}+\cgsl{2}\right)\zeta^{(\pm)}_{a}
+\frac{1}{8}\bar{\zeta}_{a}^{(\pm)}\left(i\cgtsl{3}+\cgtsl{2} \right)\gamma^a\tilde{\lambda}^{(\pm)}
\nonumber\\
&
+\frac{1}{12}i\bar{\tilde{\lambda}}^{(\pm)}\left(i\cgsl{3}+ \cgsl{2}\right)\eta^{(\pm)}
+\frac{1}{12}i\bar{\eta}^{(\pm)}\left(i\cgtsl{3}+  \cgtsl{2}\right)\tilde{\lambda}^{(\pm)} 
\nonumber\\
&
+\frac{1}{8}i\bar{\tilde{\lambda}}^{(\pm)}\Bigl(i\cgsl{3}-\cgsl{2}\Bigr)\xi^{(\pm)}
+\frac18i\bar{\xi}^{(\pm)}\left(i\cgtsl{3}-  \cgtsl{2}\right)\tilde{\lambda}^{(\pm)} 
\nonumber\\
 &
 -\frac14i\bar{\zeta}_{a}^{(\pm)}\left(\Sigma^{-2}\gamma^{[c}\Fsl_{2}\gamma^{a]} \mp 2\Sigma\gamma^{[c}\slashed{K}_2\gamma^{a]}\right)\zeta^{(\pm)}_{c} 
 \pm\Sigma\bar{\zeta}_{a}^{(\pm)}\gamma^{[c}\slashed{L}_2^{(\pm)}\gamma^{a]} \zeta^{(\mp)}_{c} 
\nonumber\\
&
 +\frac16\bar{\zeta}_{a}^{(\pm)}\left(\Sigma^{-2}\Fsl_{2} \pm\Sigma\slashed{K}_2\right)\gamma^{a}\eta^{(\pm)}
\mp \frac{1}{3}i\Sigma\bar{\zeta}_{a}^{(\pm)}\slashed{L}_2^{(\pm)}\gamma^a\eta^{(\mp)}
\nonumber\\
&
+\frac{1}{6}\bar{\eta}^{(\pm)}\gamma^{c}\left(\Sigma^{-2}\Fsl_{2}\pm\Sigma\slashed{K}_2\right)\zeta^{(\pm)}_{c}
\mp \frac{1}{3}i\Sigma \bar{\eta}^{(\pm)}\gamma^c \slashed{L}^{(\pm)}_2\zeta^{(\mp)}_{c}
\nonumber
\\
&
+\frac{1}{8}i\bar{\tilde{\lambda}}^{(\pm)}\left(\Sigma^{-2}\Fsl_{2} \pm 2\Sigma\slashed{K}_2\right)\tilde{\lambda}^{(\pm)}
\pm \frac{1}{2}\Sigma \bar{\tilde{\lambda}}^{(\pm)}\slashed{L}^{(\pm)}_2 \tilde{\lambda}^{(\mp)}
\nonumber\\
&
+\frac18i\bar{\xi}^{(\pm)}\Bigl(\Sigma^{-2}\Fsl_{2} \mp 2 \Sigma \slashed{K}_{2}\Bigr)\xi^{(\pm)}
\mp \frac12 \Sigma\bar{\xi}^{(\pm)}\slashed{L}^{(\pm)}_{2}\xi^{(\mp)}
\nonumber\\
&
-\frac{1}{36}i\bar{\eta}^{(\pm)}\left(5\Sigma^{-2}\Fsl_{2}\pm2\Sigma\slashed{K}_2\right)\eta^{(\pm)}
   \mp \frac{1}{9}\Sigma\bar{\eta}^{(\pm)}\slashed{L}^{(\pm)}_2\eta^{(\mp)}
\end{align}

\noindent Similarly, the interaction Lagrangian for the coupling to the charge conjugate fields reads
\begin{align}
\mathcal{L}_{\bar{\psi}\psi^\conj}^{(\pm)}
={}&  \mp\frac{1}{2}\bar{\tilde{\lambda}}^{(\pm)}\gamma^a \Psl {\zeta}^{(\mp)\conj}_{a} \pm \frac12\fudge\bar{\zeta}_{a}^{(\pm)} \Psl\gamma^a\tilde{\lambda}^{(\mp)\conj}
\nonumber\\
&
\pm \frac{1}{4}\bar{\zeta}_{a}^{(\pm)} \gamma^{[a}\left(-i\cgsl{3}+\cgsl{2} \pm2\cgsl{1}\right)\gamma^{d]} \zeta^{(\mp)\conj}_{d} 
+ \bar{\zeta}_{a}^{(\pm)} \left[i\cnpm{1}{}{b}\gamma^{abd} 
-
\cnpm{0}{}{}
\gamma^{ad}\right]\zeta^{(\pm)\conj}_{d}
\nonumber\\
&
\mp\frac{1}{12}i\bar{\zeta}_{a}^{(\pm)} \left(i\cgsl{3}-\cgsl{2} 
\right)\gamma^{a}\eta^{(\mp)\conj} - \frac{2}{3}i\cnpm{0}{}{}\bar{\zeta}_{a}^{(\pm)} \gamma^{a}\eta^{(\pm)\conj}
\nonumber\\
&
\mp \frac{1}{12}i\bar{\eta}^{(\pm)}\gamma^d
\left(i\cgsl{3} - \cgsl{2}
\right)\zeta^{(\mp)\conj}_{d}
-\frac{2}{3}i\cnpm{0}{}{}\bar{\eta}^{(\pm)}\gamma^{d}\zeta^{(\pm)\conj}_{d}
\nonumber\\
&  
\mp \frac18i \bar{\zeta}_{a}^{(\pm)} \left(i\cgsl{3} +\cgsl{2}\pm 2\cgsl{1}
\right)\gamma^{a}\xi^{(\mp)\conj}
+ \frac12\bar{\zeta}_{a}^{(\pm)} \left( \cnpmsl{1}
-i\cnpm{0}{}{}\right)\gamma^a\xi^{(\pm)\conj}
\nonumber\\
&
\mp \frac{1}{8}i\bar{\xi}^{(\pm)}\gamma^d\left(i\cgsl{3}+\cgsl{2}\pm 2\cgsl{1}
\right)\zeta^{(\mp)\conj}_{d}
+ \frac12\bar{\xi}^{(\pm)}\gamma^d\left(
\cnpmsl{1}-i\cnpm{0}{}{}\right) \zeta^{(\pm)\conj}_{d} 
\nonumber\\
&
\pm\frac{1}{12}\bar{\xi}^{(\pm)}\left(i\cgsl{3}+ \cgsl{2}
\right)\eta^{(\mp)\conj}
+\frac23\cnpm{0}{}{}\bar{\xi}^{(\pm)}\eta^{(\pm)\conj}
\pm\frac{3}{16}\bar{\xi}^{(\pm)}\cgsl{2}{\xi}^{(\mp)\conj}
\nonumber\\
&
\mp\frac{1}{36}\bar{\eta}^{(\pm)}\left(i\cgsl{3}-\cgsl{2}\mp 6 \cgsl{1}
\right)\eta^{(\mp)\conj}
+ \frac{1}{9}i\bar{\eta}^{(\pm)}\left(3\cnpmsl{1}-5i\cnpm{0}{}{}
\right)\eta^{(\pm)\conj} 
\nonumber\\
& 
\pm\frac{1}{12}\bar{\eta}^{(\pm)}\left( i\cgsl{3} +\cgsl{2}
\right)\xi^{(\mp)\conj}
+\frac{2}{3}\cnpm{0}{}{}\bar{\eta}^{(\pm)}\xi^{(\pm)\conj}
\end{align}

\noindent where, in a slight abuse of notation, $\Psl$ now denotes the 5-$d$ quantity $\Psl = (1/2)\gamma^{b}\left(\partial_{b}\phi + ie^{\phi}\partial_{b}a\right)$.

It is worth noticing that this action can be also obtained by direct dimensional reduction of the following $D=10$ action:
\begin{align}
S_{9+1} ={}&
 K_{10}\int d^{10}x\sqrt{-g_{10}}\biggl[\frac{1}{2}\bar{\lambda}\left(\hat{\nablasl} -\frac{3i}{2}\slashed{Q} - \frac{i}{8}\Fsl_{(5)}\right)\lambda +\frac{1}{8}\left(\bar{\Psi}_{A}\Gsl^{*}\Gamma^A\lambda - \bar{\lambda}\Gamma^A \Gsl \Psi_A\right)
\nonumber\\
&\hphantom{K\int d^{10}x\sqrt{-g_{10}}\biggl[}
-\frac{1}{4}\left(\bar{\lambda}\Gamma^{A}\Psl \Psi_{A}^\conj +\bar{\Psi}_{A}\Psl \Gamma^{A}\lambda^\conj + \frac{1}{8}\bar{\Psi}_{A}\Gamma^{ABC}S_{B}\Psi_{C}^\conj +\mbox{ c.c.}\right)
\nonumber\\
&\hphantom{K\int d^{10}x\sqrt{-g_{10}}\biggl[}
+\bar{\Psi}_{A}\Gamma^{ABC}\left(\hat{\nabla}_{B}-\frac{i}{2}Q_{B} + \frac{i}{16}\Fsl_{(5)}\Gamma_{B}\right)\Psi_{C}
\biggr],
\end{align}

\noindent from which the 10-$d$ fermionic equations of motion can be derived. As usual in the context of AdS/CFT, the bulk action would have to be supplemented by appropriate boundary terms in order to compute correlation functions of the dual field theory operators holographically. 
\section{$N=4$ supersymmetry}\label{section:N=4}

It is expected that the Lagrangian we have derived has $N=4$ $d=5$ supersymmetry, and we will provide evidence that that is the case. We expect to find the gravity multiplet (containing the graviton, the scalar $\Sigma$ and vectors) and a pair of vector multiplets (containing the rest of the scalars and vectors). Let us consider the supersymmetry variations of the 10-$d$ theory. 
These are
\beqn
\delta\lambda &=& \Psl\varepsilon^\conj+\frac{1}{4}\Gsl\varepsilon\\
\delta\Psi_A&=& \hat\nabla_A\varepsilon-\frac12iQ_A\varepsilon+\frac{i}{16}\Fsl_{(5)}\Gamma_A\varepsilon-\frac{1}{16}S_A\varepsilon^\conj
\eeqn
where
\begin{equation}
S_{A} = \frac{1}{6}\left({\Gamma_{A}}^{DEF}G_{DEF}- 9\Gamma^{DE}G_{ADE}\right)
=\Gamma_{A}\Gsl - 2G_{ADE}\Gamma^{DE}
\end{equation}

\noindent as before. Given the consistent truncation (assuming throughout that the $SE_5$ is not $S^5$), the variational parameters must also be $SU(2)$
singlets:
\beqn
\varepsilon&=&e^{W/2}\theta^{(+)}(x)\otimes \varepsilon_+(y)e^{\frac32 i\chi}\otimes u_-
+e^{W/2}\theta^{(-)}(x)\otimes \varepsilon_-(y)e^{-\frac32 i\chi}\otimes u_-\\
\varepsilon^\conj&=&e^{W/2}\theta^{(-)\conj}(x)\otimes \varepsilon_+(y)e^{\frac32 i\chi}\otimes u_-
-e^{W/2}\theta^{(+)\conj}(x)\otimes \varepsilon_-(y)e^{-\frac32 i\chi}\otimes u_-\, .
\eeqn
The evaluation of the variations proceeds much as the calculations leading to the equations of motion, and we find
\beqn
\delta\tilde\lambda^{(\pm)}&=& 
\pm \Psl\theta^{(\mp)\conj}
-
\frac14\Bigl(i\cgsl{3}+ \cgsl{2}\mp 2\cgsl{1}
\Bigr)\theta^{(\pm)}
\mp i\left(\cnpmsl{1}-i \cnpm{0}{}{}\right)\theta^{(\mp)}\\
\delta\xi^{(\pm)}
&=&
\Bigl[2i(\pasl U) + e^{-4U} \slashed{K}_1-2i e^{Z+4W}\pm2 i e^{-4U}\Sigma^{-1}\Bigr]\theta^{(\pm)} 
\nonumber\\
&&
\mp\frac{1}{4}\Bigl(\cgsl{3}-i\cgsl{2}\mp 2i\cgsl{1}
\Bigr)\theta^{(\mp)\conj}  
- \left(\cnpmsl{1}-i\cnpm{0}{}{}\right)\theta^{(\pm)\conj} 
\eeqn
\beqn
\delta \eta^{(\pm)}
&=&
\Biggl[
-\frac32i(\Sigma^{-1}\pasl\Sigma)
-\frac12\Sigma^{-2}\Fsl_{2}
\mp\frac12\Sigma\slashed{K}_2
\mp ie^{-4U}\Sigma^{-1}
\pm 3i\Sigma^2
-2ie^{Z+4W}
\Biggr]\theta^{(\pm)}
\nonumber
\\&&
\pm i\Sigma\slashed{L}_2^{(\pm)}\theta^{(\mp)}
\mp\frac{1}{4}\Bigl(\cgsl{3}+i\cgsl{2}
\Bigr)\theta^{(\mp)\conj}
+2i\cnpm{0}{}{}\theta^{(\pm)\conj}\\
\delta\zeta_a^{(\pm)}
&=&
\left[
\nabla_a\mp \frac32iA_a + \frac{1}{4}ie^{\phi}\partial_{a} a-\frac{1}{2}ie^{-4U}K_{1a}\right]\theta^{(\pm)}
+\gamma_a
\Biggl(
\pm \frac{1}{3}e^{-4U}\Sigma^{-1}
\pm\frac12 \Sigma^2
-\frac13e^{Z+4W}
\Biggr)\theta^{(\pm)}
\nonumber\\
&&
+\frac18i\Sigma^{-2}\left(\slashed{F}_{2} \gamma_a-\frac13\gamma_a\slashed{F}_2\right) \theta^{(\pm)}
\mp\frac14i\Sigma\left(\slashed{K}_{2} \gamma_a-\frac13\gamma_a\slashed{K}_2\right) \theta^{(\pm)}
\nonumber\\
&&
\mp\frac{1}{8}\Biggl[ 
i\left(\cgsl{3} \gamma_a-\frac13\gamma_a\cgsl{3}\right)
-\left(\cgsl{2} \gamma_a-\frac13\gamma_a\cgsl{2}\right)
\mp 4\cg{1}{}{a}
\Biggr] \theta^{(\mp)\conj}
\nonumber\\
&&
\mp\frac12\Sigma\left(\slashed{L}_{2}^{(\pm)} \gamma_a-\frac13\gamma_a\slashed{L}_2^{(\pm)}\right) \theta^{(\mp)}
+ \Bigl(
i\cnpm{1}{}{a}
+\frac{1}{3}\cnpm{0}{}{} \gamma_a
\Bigr)
\theta^{(\pm)\conj}\, .
\eeqn

Consulting for example \cite{Schon:2006kz, DallAgata:2001vb}, one sees immediately that it is $\delta\eta^{(\pm)}$ that contains $\Sigma^{-1}\pasl\Sigma$, and thus we deduce that it is $\eta^{(\pm)}$ that sits in the $N=4$ gravity multiplet.
These could be assembled into four symplectic-Majorana spinors, forming the ${\bf 4}$ of $USp(4)\sim SO(5)$. The remaining fermions $\xi^{(\pm)},\tilde\lambda^{(\pm)}$ can then be arranged into an $SO(2)$ doublet of $USp(4)$ quartets, appropriate to the pair of vector multiplets.

\section{Linearized analysis}\label{section:vacua}

\subsection{The supersymmetric vacuum solution}

It has been shown that the $N=4$ possesses a supersymmetric vacuum with $N=2$ supersymmetry. To see the details of the St\"uckelberg mechanism at work, we linearize the fermions around the vacuum, in which all of the fluxes are zero and  the scalars take the values  $U=V=X=Y=Z=0$. Around this vacuum, the supersymmetry variations reduce to
\beqn
\delta \eta^{(+)}
&=&\delta\xi^{(+)}
=\delta\tilde\lambda^{(+)}
=0
\\
\delta\zeta_a^{(+)}
&=&
D_a\theta^{(+)}+\frac12\gamma_a\theta^{(+)}
\\
\delta \eta^{(-)}
&=&
\delta\xi^{(-)}
=
-4i  \theta^{(-)}
\\
\delta\lambda^{(-)}
&=&
0
\\
\delta\zeta_a^{(-)}
&=&
D_a\theta^{(-)}-\frac76\gamma_a\theta^{(-)}\, .
\eeqn
These correspond to unbroken $N=2$ supersymmetry parametrized by $\theta^{(+)}$, while the supersymmetry given by $\theta^{(-)}$ is broken.
In our somewhat unusual normalizations of the fermions, as given in (\ref{fermionnormsone}), we can deduce that the Goldstino is
proportional to $g=\frac{1}{10}\left(\eta^{(-)}+\frac32\xi^{(-)}\right)$ (orthogonal to the invariant mode $\frac{1}{10}\left(\eta^{(-)}-\xi^{(-)}\right)$).
The kinetic terms in this vacuum then take the form
\begin{align}
S_{svac}={}&
\frac12\left(\bar{\tilde\lambda}^{(+)}\Dsl\ \tilde\lambda^{(+)}-\frac72\bar{\tilde\lambda}^{(+)} \tilde\lambda^{(+)}\right)
+\frac12\left(\bar{\tilde\lambda}^{(-)}\Dsl\ \tilde\lambda^{(-)}+\frac32\bar{\tilde\lambda}^{(-)} \tilde\lambda^{(-)}\right)
\nonumber
\\
&
+\frac{2}{15}\left(\bar\kappa_1^{(+)}\Dsl\ \kappa_1^{(+)}-\frac{11}{2}\bar\kappa_1^{(+)}\kappa_1^{(+)}\right)
+\frac15\left(\bar\kappa_2^{(+)}\Dsl\ \kappa_2^{(+)}+\frac92\bar\kappa_2^{(+)}\kappa_2^{(+)}\right)
 +20\left(\bar{h}\Dsl\ h-\frac52\bar hh\right)
\nonumber
\\
&
+\bar{\zeta}^{(-)}_a\gamma^{abc}D_b\zeta^{(-)}_{c} 
+\frac72\bar{\zeta}^{(-)}_a\gamma^{ac}\zeta^{(-)}_{c}
+\left(\frac{40}{3}i\bar{\zeta}^{(-)}_a\gamma^{a}g
+\mbox{c.c.}\right)
-\frac{700}{9}\bar{g}g
+\frac{40}{3}\bar{g}\Dsl\ g
\nonumber
\\
&
+\bar{\zeta}^{(+)}_a\gamma^{abc}D_b\zeta^{(+)}_{c} 
-\frac32\bar{\zeta}^{(+)}_a\gamma^{ac}\zeta^{(+)}_{c}\, ,
\end{align}
where $\kappa_{1,2}^{(+)}$ are linear combinations of $\eta^{(+)},\xi^{(+)}$. Since the geometry is $AdS_5$, the fourth line represents a ``massless" gravitino, while, defining the invariant  combination $\Psi_a=\zeta_a^{(-)}+\frac76i\gamma_ag-iD_ag$, the third line becomes
\beq
\bar{\Psi}_a\gamma^{abc}D_b\Psi_c+\frac72\bar{\Psi}_a\gamma^{ab}\Psi_b\, ,
\eeq
the action of a massive gravitino. This is the Proca/St\"uckelberg mechanism. We see then that we have fermion modes of mass $\{\frac{11}{2},\frac72,\frac52,\frac32,-\frac32,-\frac72,-\frac92\}$ which correspond to the fermionic modes of unitary irreps of $SU(2,2|1)$ and which also coincide with the lowest rungs of the KK towers of the sphere compactification \cite{Kim:1985ez}. The corresponding features in the bosonic spectrum were noted in \cite{Cassani:2010uw,Gauntlett:2010vu}. Specifically, in the language of Ref. \cite{Gunaydin:1984fk}, the $p=2$ sector contains $\zeta_a^{(+)},\tilde\lambda^{(-)}$, $p=3$ contains $\zeta_a^{(-)},\tilde\lambda^{(+)},\eta^{(-)},\xi^{(-)}$ and $p=4$ contains $\eta^{(+)},\xi^{(+)}$.

\subsection{The Romans $AdS_{5}$ vacuum}
The non-supersymmetric $AdS$ vacuum \cite{Romans:1984an,Gunaydin:1984fk} of the theory has radius $\sqrt{8/9}$, and vevs
\begin{equation}
e^{4U} = e^{-4V}=\frac{2}{3}\, , \qquad Y = \frac{e^{i\theta}}{\sqrt{12}}e^{\phi/2}\, \qquad X =(a+ ie^{-\phi})Y\, ,
\end{equation}
where $\theta$ is an arbitrary constant phase. The axion $a$ and dilaton $\phi$ are arbitrary \cite{Cassani:2010uw,Gauntlett:2010vu}.
For the various quantities appearing in the effective action we have
\begin{equation}
\mathcal{G}_{i} = \tilde{\mathcal{G}}_{i} = \mathcal{N}_{1}^{(\pm)}=\tilde{\mathcal{N}}_{1}^{(\pm)}=\mathcal{N}_{0}^{(+)}=\tilde{\mathcal{N}}_{0}^{(-)}=K_{1}=K_{2}=L_{2}=0\, ,
\end{equation}

\noindent where $i=1,2,3$, and 
\begin{equation}
e^{-4W} = \frac{2}{3}\,,\quad\Sigma = 1\,, \quad e^{Z}=\frac{1}{2}\,,\quad P=0\,, \quad \left(\cnm{0}{}{}\right)^{*}=\cntp{0}{}{} =-\frac{3}{\sqrt{2}}e^{i\theta}\,.
\end{equation}
\noindent We then find
\begin{align}
\mathcal{L}_{mass}^{(+)}
 ={}&
 - \frac{15}{8}\bar{\tilde{\lambda}}^{(+)}\tilde{\lambda}^{(+)}
-\frac{9}{4}\bar{\zeta}_{a}^{(+)}\gamma^{ac}\zeta^{(+)}_{c}
+\frac{1}{4}\bar{\eta}^{(+)}\eta^{(+)}
+ \frac{3}{8}\bar{\xi}^{(+)}\xi^{(+)}
\nonumber\\
&
- 2
\left(\bar{\eta}^{(+)}\xi^{(+)}+\bar{\xi}^{(+)}\eta^{(+)}\right)
- \frac{3}{4}i
\left(\bar{\zeta}_{a}^{(+)}\gamma^{a}\xi^{(+)}+\bar{\xi}^{(+)}\gamma^a\zeta^{(+)}_{a} \right)
\nonumber\\
&
 -\frac{3}{\sqrt{2}}e^{i\theta}\left(
 \frac12\bar{\zeta}_{a}^{(+)}\gamma^a\tilde{\lambda}^{(-)}
  +\frac{2}{3}i \bar{\eta}^{(+)}\tilde{\lambda}^{(-)}
+\frac12i\bar{\xi}^{(+)}\tilde{\lambda}^{(-)}
 \right)
\end{align}
\begin{align}
\mathcal{L}_{mass}^{(-)}
 ={}&
 \frac{9}{8}\bar{\tilde{\lambda}}^{(-)}\tilde{\lambda}^{(-)}
+\frac{15}{4}\bar{\zeta}_{a}^{(-)}\gamma^{ac}\zeta^{(-)}_{c}
-\frac{13}{12}\bar{\eta}^{(-)}\eta^{(-)}
- \frac{21}{8}\bar{\xi}^{(-)}\xi^{(-)}
\nonumber\\
&
+i\left(\bar{\zeta}_{a}^{(-)}\gamma^{a}\eta^{(-)}+\bar{\eta}^{(-)}\gamma^{a}\zeta^{(-)}_{a}\right)
+ \frac{9}{4}i\left(\bar{\zeta}_{a}^{(-)}\gamma^{a}\xi^{(-)}+\bar{\xi}^{(-)}\gamma^a\zeta^{(-)}_{a} \right)
\nonumber\\
&
+\frac{3}{\sqrt{2}}e^{-i\theta}\left(
\frac12\bar{\tilde{\lambda}}^{(-)}\gamma^a\zeta_a^{(+)}
+\frac23 i\bar{\tilde{\lambda}}^{(-)}\eta^{(+)}
+\frac12i\bar{\tilde{\lambda}}^{(-)}\xi^{(+)}
\right)
\end{align}
\begin{align}
\mathcal{L}_{\bar{\psi}\psi^\conj}^{(-)}
={}& 
\frac{3}{\sqrt{2}}e^{-i\theta}\biggl(\bar{\zeta}_{a}^{(-)} \gamma^{ad}\zeta^{(-)\conj}_{d} 
-\frac{5}{9}\bar{\eta}^{(-)}\eta^{(-)\conj} 
+ \frac{2}{3}i\bar{\zeta}_{a}^{(-)} \gamma^{a}\eta^{(-)\conj}
+\frac{2}{3}i\bar{\eta}^{(-)}\gamma^{d}\zeta^{(-)\conj}_{d}
\nonumber\\
&
\hphantom{\cnpm{0}{}{}\Bigl(}
+\frac{i}{2}\bar{\zeta}_{a}^{(-)} \gamma^a\xi^{(-)\conj}
+\frac{i}{2}\bar{\xi}^{(-)}\gamma^d\zeta^{(-)\conj}_{d} 
- \frac23\bar{\xi}^{(-)}\eta^{(-)\conj}
-\frac{2}{3}\bar{\eta}^{(-)}\xi^{(-)\conj}\biggr)
\end{align}

\noindent and
\begin{align}
\mathcal{L}_{1}^{(\pm)}=\mathcal{L}_{2}^{(\pm)}=\mathcal{L}_{\bar{\psi}\psi^\conj}^{(+)}
 ={}&
 0\, .
\end{align}
We see by inspection that indeed both gravitinos are massive. For example, $\zeta_a^{(+)}$ eats the goldstino proportional to  $g^{(+)}=\frac32 i\xi^{(+)}-{\cnm{0}{}{}}^*\tilde\lambda^{(-)}$, while the Goldstino eaten by $\zeta_a^{(-)}$ is a linear combination of $\xi^{(-)},\eta^{(-)}$ and their conjugates.

\section{Examples}\label{section:Examples}
As an application of our general result \eqref{effective 5d action}, in this section we discuss the coupling of the fermions to some further bosonic truncations of interest, including the minimal gauged $N=2$ supergravity theory in $d=5$, and the holographic $AdS_{5}$ superconductor of \cite{Gubser:2009qm}.

\subsection{Minimal $N=2$ gauged supergravity in five dimensions}
Perhaps the simplest further truncation one could consider that retains fermion modes entails taking $U=V=Z=K_{1}=L_{2}=G_{i}=H_{i}=M_{q}=N_{q}=0$ ($i=1,2,3$ and $q=0,1$) and $K_{2}=-F_{2}$.  It is then consistent to set $\tilde{\lambda}^{(\pm)}=\eta^{(\pm)} = \xi^{(\pm)}=0$ together with $\zeta_{a}^{(-)}=0$. This gives the right fermion content of minimal $N=2$ gauged supergravity in $d=5$, which is one Dirac gravitino ($\zeta_{a}^{(+)}$ in our notation), with an action given by
\begin{align}
S_{4+1} = K_{5}\int d^{5}x\sqrt{-g_{5}^E}\biggl[&
 \bar{\zeta}^{(+)}_{a}\gamma^{abc}D_{b}\zeta_{c}^{(+)} 
+ \mathcal{L}_{\bar{\psi}\psi}^{(+)}\biggr]
\end{align}

\noindent where
\begin{align}
\mathcal{L}_{\bar{\psi}\psi}^{(+)}
 ={}& -\frac{3}{2}\bar{\zeta}_{a}^{(+)}\gamma^{ac}\zeta^{(+)}_{c} -\frac{3}{4}i\bar{\zeta}_{a}^{(+)}\gamma^{[c}\Fsl_{2}\gamma^{a]}\zeta^{(+)}_{c}\, ,
\end{align}

\noindent and $D_{a} = \nabla_{a} -(3i/2)A_{1a}$ as before.

\subsection{No $p=3$ sector}
\begin{figure}[h!]
\begin{center}
\includegraphics[width=4.0in]{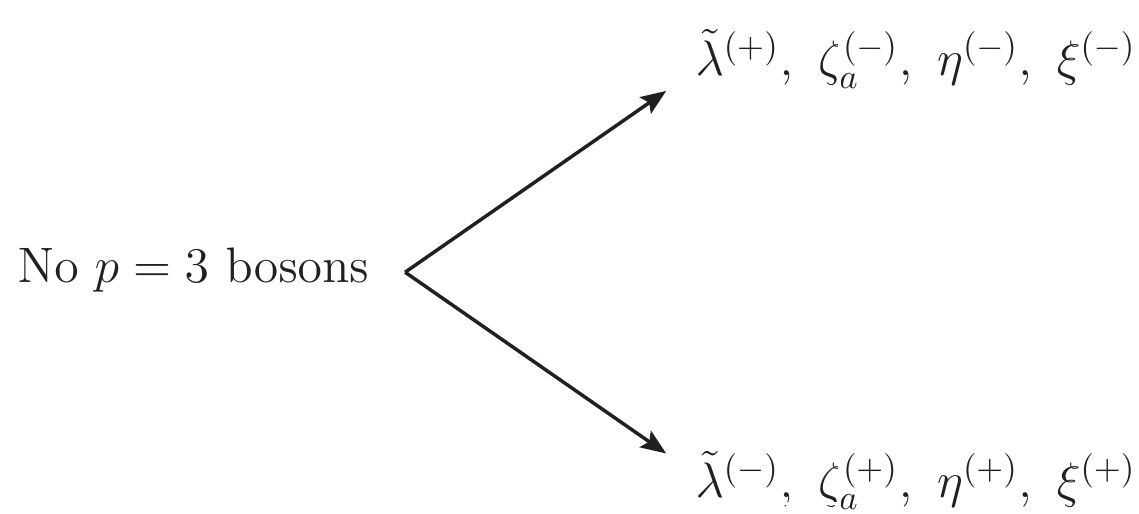}
\caption[FIG. \arabic{figure}.]{\small{Decoupling of the fermion modes in the futher truncation obtained by eliminating the bosons in the ``$ p=3$ sector".}}
\label{nop3}
\end{center}
\end{figure}

A possible further truncation of the bosonic sector considered in \cite{Gauntlett:2010vu} entails taking $G_{i}=H_{i}=L_{2}=0$ ($i = 1,2,3$). In the notation of \cite{Gunaydin:1984fk}, this corresponds to eliminating the bosonic fields belonging to the $p=3$ sector. By studying the equations of motion provided in appendix \ref{Appendix:EOM} we find that the fermion modes split into two decoupled sectors, as depicted in figure \ref{nop3}. It is therefore consistent to set the modes in either of these sectors to zero. 

We note the first set of fermion fields are all in the $p=3$ sector, while the second set are in $p=2,4$. It seems reasonable therefore to suggest that the latter truncation corresponds to an  $N=2$ gauged supergravity theory coupled to a vector multiplet and two hypermultiplets (this was suggested in \cite{Cassani:2010uw,Gauntlett:2010vu} in the context of the bosonic sector.) The former truncation would apparently be non-supersymmetric.

\subsection{Type IIB holographic superconductor}
As discussed in \cite{Cassani:2010uw,Gauntlett:2010vu}, the type IIB holographic superconductor of \cite{Gubser:2009qm} can be obtained by truncating out the bosons of the $p=3$ sector as discussed above, and further setting $a=\phi=h=0$ and $X = iY$, $K_{2}=-F_{2}$, $e^{4U} = e^{-4V} = 1 - 4|Y|^{2}$, which implies $\tilde{E}_{1}=0$ and
\begin{equation}
e^{Z} = 1 - 6|Y|^{2}\, ,\qquad K_{1} = 2i\left(Y^* DY - Y DY^*\right)\equiv 2iY^*\overleftrightarrow{D}Y\, .
\end{equation}

\noindent In terms of the variables we have defined, this truncation implies 
\begin{equation}
\mathcal{G}_{i}=\tilde{\mathcal{G}}_{i}= \mathcal{N}_{q}^{(+)}=\tilde{\mathcal{N}}_{q}^{(-)}=0
\end{equation}

\begin{figure}[h!]
\begin{center}
\includegraphics[width=6.2in]{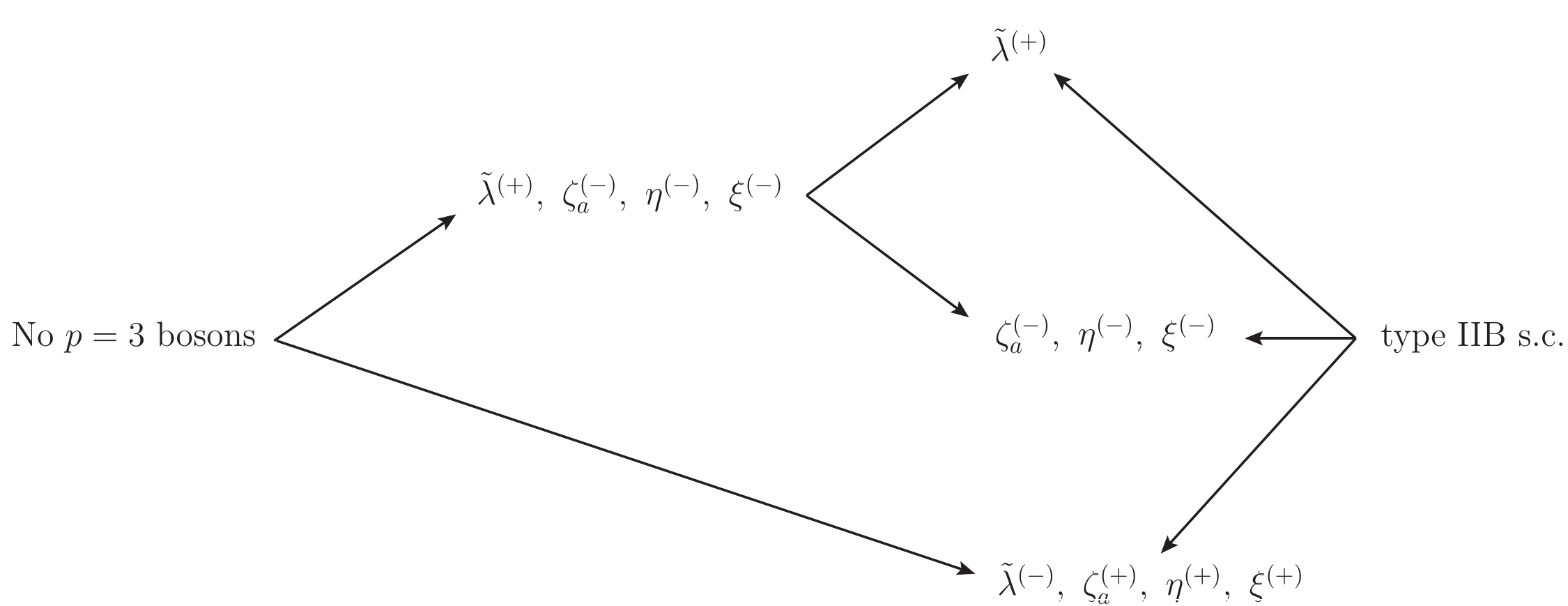}
\caption[FIG. \arabic{figure}.]{\small{Further decoupling of fermion modes in the type IIB holographic superconductor truncation.}}
  \label{typeIIBsc}
\end{center}
\end{figure}

\noindent ($i=1,2,3$ and $q=0,1$) together with
\begin{align}
\mathcal{N}_{1}^{(-)} &= -2ie^{-2U}DY^{*}\, ,& \mathcal{N}_{0}^{(-)} &= -6e^{-2U}Y^{*}\, ,
\nonumber\\
\tilde{\mathcal{N}}_{1}^{(+)} &= 2ie^{-2U}DY\, ,&  \tilde{\mathcal{N}}_{0}^{(+)} &= -6e^{-2U}Y\,,
\end{align}

\noindent and  
\begin{equation}
P=0\, ,\qquad \Sigma = 1\, , \qquad e^{-4W} = 1 - 4|Y|^{2}\, .
\end{equation}
\noindent By analyzing the equations of motion given in appendix \ref{Appendix:EOM}, we find that in this case there is a further decoupling of the fermion modes with respect to the no $p=3$ sector truncation discussed above. As depicted in figure \ref{typeIIBsc}, the $\tilde{\lambda}^{(+)}$ mode now decouples from $\zeta_{a}^{(-)}$, $\eta^{(-)}$, $\xi^{(-)}$ as well, resulting in three fermion sectors, which can then be set to zero independently. 

\subsubsection{A single spin-1/2 fermion}
The simplest scenario corresponds of course to keeping the $\tilde{\lambda}^{(+)}$ mode only, for which the effective action \eqref{effective 5d action} reduces to
\begin{align}\label{effective 5d action just lambda}
S_{4+1} = K_{5}\int d^{5}x\sqrt{-g_{5}^E}\biggl[&\frac{1}{2}\bar{\tilde{\lambda}}^{(+)}\Dsl \tilde{\lambda}^{(+)} + \mathcal{L}_{\bar{\psi}\psi}^{(+)}\biggr]
\end{align}

\noindent with
\begin{align}
\mathcal{L}_{\bar{\psi}\psi}^{(+)}
 ={}& 
 -\frac{1}{2}\bar{\tilde{\lambda}}^{(+)}\left(\frac{3}{2} +\frac{1}{4}i\Fsl_{2}
 +\frac{2-6|Y|^{2} + Y^*\overleftrightarrow{\Dsl}\,Y}{1-4|Y|^{2}}\right)\tilde{\lambda}^{(+)} \, ,
\end{align}

\noindent where we recall that $DY = dY - 3iA_{1}Y$, and $\Dsl\, \tilde{\lambda}^{(+)} = \left(\nablasl -\frac{3i}{2}\slashed{A}_{1}\right)\tilde{\lambda}^{(+)}$. As pointed out in \cite{Gauntlett:2010vu}, we can make contact with the notation of \cite{Gubser:2009qm} by setting $A_{1} = (2/3)A$ and $Y=(1/2)e^{i\theta}\tanh (\eta/2)$. Notice that $\tilde{\lambda}^{(+)}$ only couples derivatively to the phase of the charged scalar $Y$. The model \eqref{effective 5d action just lambda} is particularly well suited for an exploration of fermion correlators via holography, inasmuch as the presence of a single spin-1/2 field makes the application of all the standard gauge/gravity duality techniques possible.  Naturally, such a program becomes more involved in the presence of mixing between the gravitino and the spin-1/2 fields. 

\subsubsection{Retaining half of the fermionic degrees of freedom}
For the $\tilde{\lambda}^{(-)}, \zeta_{a}^{(+)},\xi^{(+)}, \eta^{(+)}$ sector we find that \eqref{effective 5d action} reads
\begin{align}\label{effective 5d action v2}
S_{4+1} = K_{5}\int d^{5}x\sqrt{-g_{5}^E}\biggl[&\frac{1}{2}\bar{\tilde{\lambda}}^{(-)}\Dsl \tilde{\lambda}^{(-)} + \bar{\zeta}^{(+)}_{a}\gamma^{abc}D_{b}\zeta_{c}^{(+)} 
+ \frac{1}{3}\bar{\eta}^{(+)}\Dsl\, \eta^{(+)}
\nonumber\\
&+\frac12\bar{\xi}^{(+)}\Dsl\, \xi^{(+)} + \mathcal{L}_{\bar{\psi}\psi}
\biggr]
\end{align}
\noindent with 
\begin{align}
\mathcal{L}_{\bar{\psi}\psi}
={}&
\frac{3}{8}i\bar{\tilde{\lambda}}^{(-)}\Fsl_{2}\tilde{\lambda}^{(-)}
+ 3\left(e^{-4U}|Y |^{2}+\frac{1}{4}\right)\bar{\tilde{\lambda}}^{(-)}\tilde{\lambda}^{(-)} -\frac{1}{2} e^{-4U}\bar{\tilde{\lambda}}^{(-)} \Bigl(Y^*\overleftrightarrow{\Dsl}\,Y\Bigr)\tilde{\lambda}^{(-)} 
\nonumber\\
& 
-\frac{3}{4}i \bar{\zeta}_{a}^{(+)}\gamma^{[c}\Fsl_{2}\gamma^{a]}\zeta^{(+)}_{c} 
-3\left(2e^{-4U}|Y|^{2}+\frac{1}{2}\right)\bar{\zeta}_{a}^{(+)}\gamma^{ac}\zeta^{(+)}_{c}
- e^{-4U} \bar{\zeta}_{a}^{(+)}\gamma^{[c}\Bigl(Y^*\overleftrightarrow{\Dsl}\,Y\Bigr)\gamma^{a]}
 \zeta^{(+)}_{c}
\nonumber\\
&
-\frac{i}{12}\bar{\eta}^{(+)}\Fsl_{2}\eta^{(+)}
+\frac{1}{6}e^{-4U}\bar{\eta}^{(+)}\left(
1 + 2Y^*\overleftrightarrow{\Dsl}\,Y\right)\eta^{(+)}
\nonumber\\
&
+ \frac{3}{8}i\bar{\xi}^{(+)}\Fsl_{2} \xi^{(+)}+ \frac{3}{4}e^{-4U}\bar{\xi}^{(+)}\left(3 -2Y^*\overleftrightarrow{\Dsl}\,Y\right)\xi^{(+)}
-3\bar{\xi}^{(+)}\xi^{(+)}
\nonumber
\\
&
-e^{-2U}\bar{\tilde{\lambda}}^{(-)}\gamma^a \Bigl(\Dsl\,Y^{*}
-3Y^{*}\Bigr)\zeta^{(+)}_{a} 
-e^{-2U}\bar{\zeta}_{a}^{(+)}\Bigl(\Dsl\,Y
+3e^{-2U}Y\Bigr)\gamma^a\tilde{\lambda}^{(-)}
\nonumber
\\
&
+2ie^{-4U}\bar{\xi}^{(+)}\gamma^{a}\Bigl(Y\Dsl\,Y^* - 3|Y|^{2}\Bigr)\zeta_{a}^{(+)}
-2ie^{-4U}\bar{\zeta}_{a}^{(+)}\Bigl(Y^*\Dsl\, Y +3|Y|^{2}\Bigr)\gamma^{a}\xi^{(+)}
 \nonumber
\\
& 
  +ie^{-2U}\bar{\tilde{\lambda}}^{(-)}\Bigl(\Dsl\,Y^{*}
+3Y^{*}\Bigr)\xi^{(+)}+ie^{-2U}\bar{\xi}^{(+)}\Bigl(\Dsl\,Y
-3Y\Bigr)\tilde{\lambda}^{(-)}
\nonumber
\\
&
-4i e^{-2U}\left(Y\bar{\eta}^{(+)}\tilde{\lambda}^{(-)}
-Y^{*}\bar{\tilde{\lambda}}^{(-)}\eta^{(+)}\right)
-2\left(\bar{\xi}^{(+)}\eta^{(+)}+\bar{\eta}^{(+)}\xi^{(+)}\right),
\end{align}

\noindent where we recall that $e^{4U}=1 - 4|Y|^{2}$. We note the presence of a variety of couplings between the fermions and the charged scalar, as well as Pauli couplings.

\subsubsection{The $\zeta_{a}^{(-)}$, $\eta^{(-)}$, $\xi^{(-)}$ sector}
For the remaining decoupled sector containing the $\zeta_{a}^{(-)}$, $\eta^{(-)}$, $\xi^{(-)}$ modes we find
\begin{align}\label{effective 5d action v3}
S_{4+1} = K_{5}\int d^{5}x\sqrt{-g_{5}^E}\biggl[& \bar{\zeta}^{(-)}_{a}\gamma^{abc}D_{b}\zeta_{c}^{(-)} 
+ \frac{1}{3}\bar{\eta}^{(-)}\Dsl\, \eta^{(-)}+\frac12\bar{\xi}^{(-)}\Dsl\, \xi^{(-)} 
\nonumber\\
&+ \mathcal{L}_{\bar{\psi}\psi} + \frac{1}{2}\left(\mathcal{L}_{\bar{\psi}\psi^\conj}^{(-)}+\mbox{c.c.}\right)
\biggr]
\end{align}

\noindent where now
\begin{align}
\mathcal{L}_{\bar{\psi}\psi}
 ={}&
e^{-4U}\biggl[\left(\frac{7}{2}  - 12|Y|^{2}\right)\bar{\zeta}_{a}^{(-)}\gamma^{ac}\zeta^{(-)}_{c}+\frac{1}{9}\left(-\frac{23}{2} + 60|Y|^{2}\right)\bar{\eta}^{(-)}\eta^{(-)}
\nonumber\\
&
\hphantom{e^{-4U}\biggl[}
-\frac{3}{2}\left(\frac{3}{2}  - 4|Y|^{2}\right)\bar{\xi}^{(-)}\xi^{(-)} + \frac{ 2}{3}\left(-1  +12|Y|^{2}\right)
\left(\bar{\eta}^{(-)}\xi^{(-)}+\bar{\xi}^{(-)}\eta^{(-)}\right)
\nonumber\\
&
\hphantom{e^{-4U}\biggl[}
+\frac{4}{3}i\left(1 -6|Y|^{2} \right)
\left(\bar{\zeta}_{a}^{(-)}\gamma^{a}\eta^{(-)}+\bar{\eta}^{(-)}\gamma^{a}\zeta^{(-)}_{a}\right)
\nonumber\\
&
\hphantom{e^{-4U}\biggl[}
+ 2i\left( 1- 3|Y|^{2}\right)
\left(\bar{\zeta}_{a}^{(-)}\gamma^{a}\xi^{(-)}+\bar{\xi}^{(-)}\gamma^a\zeta^{(-)}_{a} \right)
\nonumber
\\
&
\hphantom{e^{-4U}\biggl[}
-\bar{\zeta}_{a}^{(-)}
\gamma^{[c}Y^*\overleftrightarrow{\Dsl}Y \gamma^{a]}
 \zeta^{(-)}_{c}
  -\frac{3}{2}\bar{\xi}^{(-)}Y^*\overleftrightarrow{\Dsl}Y \xi^{(-)}
+\frac{1}{3}\bar{\eta}^{(-)}Y^*\overleftrightarrow{\Dsl}Y \eta^{(-)} 
\nonumber
\\
& 
\hphantom{e^{-4U}\biggl(}
-2i\bar{\zeta}_{a}^{(-)}Y^*\Dsl Y\gamma^{a}\xi^{(-)}
+2i\bar{\xi}^{(-)}\gamma^{a}Y\Dsl Y^{*}\zeta^{(-)}_{a} \biggr]
\nonumber
\\
&
+\frac14i\bar{\zeta}_{a}^{(-)}\gamma^{[c}\Fsl_{2}\gamma^{a]}\zeta^{(-)}_{c} 
-\frac18i\bar{\xi}^{(-)}\Fsl_{2}\xi^{(-)}
-\frac{7}{36}i\bar{\eta}^{(-)}\Fsl_{2}\eta^{(-)}
\nonumber\\
&
 +\frac{1}{3}\bar{\zeta}_{a}^{(-)}\Fsl_{2}\gamma^{a}\eta^{(-)}
 +\frac{1}{3}\bar{\eta}^{(-)}\gamma^{c}\Fsl_{2}\zeta^{(-)}_{c}
\end{align}

\noindent and
\begin{align}
\mathcal{L}_{\bar{\psi}\psi^\conj}^{(-)}
={}&  
 e^{-2U}\biggl[2\bar{\zeta}_{a}^{(-)} \left(\gamma^{abd} D_{b}Y^{*}
+3\gamma^{ad}Y^{*}\right)\zeta^{(-)\conj}_{d}
+4iY^{*}\left(\bar{\zeta}_{a}^{(-)} \gamma^{a}\eta^{(-)\conj} +\bar{\eta}^{(-)}\gamma^{a}\zeta^{(-)\conj}_{a}\right)
\nonumber\\
&  
\hphantom{e^{-2U}\biggl[}
- i \bar{\zeta}_{a}^{(-)} \left(\Dsl Y^* - 3Y^*\right)\gamma^a\xi^{(-)\conj}
- i\bar{\xi}^{(-)}\gamma^d\left(\Dsl Y^* - 3Y^*\right) \zeta^{(-)\conj}_{d} 
\nonumber\\
&
\hphantom{e^{-2U}\biggl[}
-4Y^*\left(\bar{\xi}^{(-)}\eta^{(-)\conj}+\bar{\eta}^{(-)}\xi^{(-)\conj}\right)
+ \frac{2}{3}\bar{\eta}^{(-)}\left(\Dsl Y^* - 5Y^*\right)\eta^{(-)\conj} 
\biggr].
\end{align}

The models \eqref{effective 5d action just lambda}, \eqref{effective 5d action v2} and \eqref{effective 5d action v3} display a variety of couplings between the fermions and the charged scalar, the fermions and their charge conjugates, and Pauli couplings as well. From the gauge/gravity duality point of view, these couplings might be of phenomenological interest and give rise to features that have not been observed so far in the simpler non-interacting fermion models in the literature. The exploration of these directions in the context of AdS/CFT will be pursued elsewhere. 

\section{Conclusions}\label{section:Conclusions}
Continuing with the program initiated in \cite{Bah:2010yt}, where we performed the reduction of the fermionic sector in the consistent truncations of $D=11$ supergravity on squashed Sasaki-Einstein seven-manifolds \cite{Gauntlett:2009zw}, in the present paper we have considered the reduction of fermions in the recently found consistent truncations of type IIB supergravity on squashed Sasaki-Einstein five-manifolds \cite{Cassani:2010uw,Liu:2010sa,Gauntlett:2010vu}. A common denominator of these KK reductions is that they consistently retain charged (massive) scalar and $p$-form fields. This feature not only establishes them as relevant from a supergravity perspective, but it also makes them particulary suitable for the description of various phenomena, such as superfluidity and superconductivity, by means of holographic techniques.

 In particular, as an application of our results we have discussed the coupling of fermions to the $(4+1)$-dimensional type IIB holographic superconductor of \cite{Gubser:2009qm}, which complements our previous result for the coupling of fermions to the $(3+1)$-dimensional M-theory holographic superconductor constructed in \cite{Gauntlett:2009dn}. It is interesting to note the differences between these two effective theories. For example, the coupling of the fermions to their charge conjugates (i.e. Majorana-like couplings) was found to play a central role in the $(3+1)$-model of \cite{Bah:2010yt}. Although such couplings are still present in the general truncation discussed in the present work, they are absent in the further truncation corresponding to the holographic $(4+1)$-dimensional superconductor. More importantly, while a simple further truncation of the fermion sector that could result in a more manageable system well suited for holographic applications eluded us in our previous work, in the present scenario we have found a very simple model (c.f. \eqref{effective 5d action just lambda}) describing a single spin-1/2 Dirac fermion interacting with the charged scalar that has been shown to condense for low enough temperatures of a corresponding black hole solution of the bosonic field equations \cite{Gubser:2009qm}. It would be interesting to apply our results to the holographic computation of fermion correlators in the presence of these superconducting condensates. Similarly, our results can be used to explore fermion correlators in other situations as well.

\vskip 1cm
\centerline{\bf Acknowledgments}

We are grateful to Leo Pando-Zayas for many helpful discussions and collaboration on an early stage of this project, and to Sean Hartnoll for helpful correspondence.  J.I.J. and R.G.L. are thankful to the Michigan Center for Theoretical Physics (MCTP) for their hospitality during the initial stages of this project. R.G.L. is supported by DOE grant FG02-91-ER40709. J.I.J. and A.T.F. are supported by  Fulbright-CONICYT fellowships. I.B. is partially supported by DOE grant DE-FG02-95ER40899 and a University of Michigan Rackham Science Award.

\appendix
\section{Conventions and useful formulae}\label{Appendix:Conventions}
In this Appendix we introduce the various conventions used in the body of the paper, and collect some useful results.

\subsection{Conventions for forms and Hodge duality}
We normalize all the form fields according to
\begin{align}\label{rewriting rform in the wedge basis}
    \omega &= \omega_{a_{1}\ldots a_{p}}\,e^{a_{1}}\otimes e^{a_{2}}\cdots \otimes e^{a_{p}}\nonumber\\
           &= \frac{1}{p!}\omega_{a_1\ldots a_{p}} \,e^{a_{1}}\wedge \cdots \wedge e^{a_{p}}\, .
\end{align}

\noindent Similarly, all the slashed $p$-forms are defined with the normalization
\begin{equation}
\slashed{\omega} = \frac{1}{p!}\gamma^{a_1\ldots a_p}\omega_{a_1\ldots a_p}\, .
\end{equation}

In $d$ spacetime dimensions, the Hodge dual acts on the basis of forms as
\begin{equation}
*(e^{a_1}\wedge \cdots \wedge e^{a_p}) = \frac{1}{(d-p)!}{\epsilon_{b_{1} \ldots b_{d-p}}}^{a_1\ldots a_p}\, e^{b_{1}}\wedge \cdots \wedge e^{b_{d-p}}\, ,
\end{equation}

\noindent where $\epsilon_{b_{1}\ldots b_{d-p}a_{1}\ldots a_{p}}$ are the components of the Levi-Civita \textit{tensor}. Equivalently, for the components of the Hodge dual $*\omega$ of a $p$-form $\omega$ we have
\begin{equation}\label{symbolicall meaning of the isomorphism via hodge}
	 (*\omega)_{a_{1}\ldots a_{d - p}} = \frac{1}{p!}{\epsilon_{a_{1}\ldots a_{d-p}}}^{b_{1}\ldots b_{p}}\omega_{b_{1}\ldots b_{p}}\, .
\end{equation}

\noindent In the $(4+1)$-dimensional external manifold $M$ we adopt the convention $\epsilon_{01234} = +1$ for the components of the Levi-Civita tensor in the orthonormal frame.

\subsection{Zehnbein and spin connection}
As discussed in section \ref{section:Ansatz}, the Kaluza-Klein metric ansatz of \cite{Cassani:2010uw}, \cite{Gauntlett:2010vu}, \cite{Liu:2010sa}, \cite{Skenderis:2010vz} is given by
\begin{equation}
ds^2_{10}=e^{2W(x)}ds_{E}^2(M)+e^{2U(x)}ds^2(KE)+e^{2V(x)}\bigl(d\chi + {\cal A}(y) +A_{1}(x)\bigr)^2\, ,
\end{equation}

\noindent where $W(x) = -\frac{1}{3}\left(4U(x) + V(x)\right)$ as in the body of the paper. We now introduce the ten-dimensional orthonormal frame $\hat{e}^{M}$. Denoting by $a,b,\ldots$ the tangent indices to $M$, by $\alpha,\beta,\ldots$ the tangent indices to the K\"ahler-Einstein base $KE$, and by $\f$ the index associated with the U(1) fiber direction $\chi$, our choice of zehnbein reads
\beqn
\hat e^a &=& e^{W}e^a\\
\hat e^\alpha &=& e^U e^\alpha\\
\hat e^\f &=& e^V \Bigl(d\chi+{\cal A}(y)+A_{1}(x)\Bigr),
\eeqn

\noindent where $e^{a}$ and $e^{\alpha}$ are orthonormal frames for $M$ and $KE$, respectively. The dual basis is then
\beqn
\hat e_a &=& e^{-W}\bigl(e_a-A_{1a}\pa_\chi\bigr)\\
\hat e_\alpha &=& e^{-U}\bigl( e_\alpha-{\cal A}_\alpha\pa_\chi\bigr)\\
\hat e_\f &=& e^{-V}\pa_\chi\, .
\eeqn

\noindent Denoting by ${\omega^a}_b$ the spin connection associated with $ds_{E}^2(M)$ and by ${\omega^\alpha}_\beta$ the spin connection appropriate to $ds^2(KE)$, for the ten-dimensional spin connection ${\hat{\omega}^{M}}_{~ \, N}$ we find
\beqn
{\hat\omega^\alpha}_{~a}&=& e^{U-W}(\pa_a U)e^\alpha\\
\hat\omega^\f_{~a}&=&e^{V-W}\left[\frac12 F_{2\, ab}e^b+(\pa_aV)\bigl(d\chi+{\cal A}+A_{1}\bigr)\right]\\
\hat\omega^\f_{~\alpha}&=&e^{V-U}\frac12{\cal F}_{\alpha\beta}e^\beta\\
\hat\omega^a_{~b}&=&\omega^a_{~b}-2\eta^{ac}\pa_{[c}W \eta_{b]d}e^{d}-\frac12 e^{2(V-W)}F^a_{2~b}\bigl(d\chi+{\cal A}+A_{1}\bigr)\\
\hat\omega^\alpha_{~\beta}&=&\omega^\alpha_{~\beta}-\frac12 e^{2(V-U)}{\cal F}^\alpha_{~\beta}\bigl(d\chi+{\cal A}+A_{1}\bigr),
\eeqn

\noindent where $\eta_{ab}$ is the flat metric in $(4+1)$ dimensions, $F_{2} \equiv dA_{1}$ and ${\cal F} \equiv d{\cal A} = 2J$, $J$ being the K\"ahler form on $KE$. 

\subsection{Fluxes}\label{Appendix:Conventions-Fluxes}
The ans\"atze for the form fields fields, reproduced here for convenience, is as presented in Ref. \cite{Gauntlett:2010vu} 
\begin{eqnarray} 
F_{(5)} &=&
4 e^{8W+Z} \textrm{vol}^E_5 
+e^{4(W+U)} * K_2 \wedge J+ K_1 \wedge J \wedge J\nonumber\\
&&+ \left[
2e^Z J \wedge J
-2e^{-8U} *K_1   +K_2 \wedge J \right]\wedge (\eta+A_1)\nonumber\\
 && + \left[ e^{4(W+U)} *L_2 \wedge \Omega + L_2 \wedge \Omega \wedge (\eta+A_1) +\mbox{c.c.} \right]
 \\[8pt]
F_{(3)} &=& G_3 +G_2 \wedge (\eta+A_1) +G_1 \wedge J +G_0\, J\wedge (\eta+A_1)
\nonumber\\
&&
+\Bigl[ N_1 \wedge \Omega  +N_0\,\Omega\wedge (\eta+A_1)+\mbox{c.c.}\Bigr]
\\[8pt]
 H_{(3)} &=& H_3 +H_2 \wedge (\eta+A_1) +H_1 \wedge J +H_0\,  J\wedge (\eta+A_1)
\nonumber\\
&&
+\Bigl[ M_1 \wedge \Omega  +M_0\,\Omega\wedge (\eta+A_1)+\mbox{c.c.}\Bigr]
 \end{eqnarray}

\noindent As pointed out in the body of the paper, notice that we have $G_{0}=H_{0}=0$ by virtue of the type IIB Bianchi identities. We will often use a complex basis on $T^{*}KE$. If $y$ denote real coordinates on $KE$, we define $z^{1} \equiv \frac{1}{2}(y^{1} + iy^{2})$, $z^{\bar{1}} \equiv \frac{1}{2}\left(y^{1} - iy^{2}\right)$, and similarly for $z^{2},z^{\bar{2}}$. With this normalization, the K\"ahler form $J$ and the holomorphic (2,0)-form $\Sigma_{(2,0)}$ are given by
\begin{align}
J &= 2i\sum_{\alpha = 1,2}e^{\alpha}\wedge e^{\bar{\alpha}}\, \\
\Sigma_{(2,0)} &= \frac{2^{2}}{2!}\epsilon_{\alpha\beta}\, e^\alpha\wedge e^\beta\, ,\label{defSigma}
\end{align}

\noindent where we have chosen $\epsilon_{12} = +1$. The components of $F_{(5)}$ with respect to the ten-dimensional frame $\hat{e}^{M}$ are then (in the real basis for $T^{*}KE$)
\begin{eqnarray}
 F_{(5)abcde} &=& 4 e^{Z+3W} \epsilon_{abcde} \\
F_{(5)abcd\f} &=& -2 e^{-4U-W} \epsilon_{abcd}^{\;\;\;\;\;\;\;e} K_{1;e} \\
F_{(5)a\alpha \beta \gamma \delta} &=& 6e^{-4U-W} K_{1;a} J_{[\alpha \beta} J_{\gamma \delta]} \\
F_{(5)\alpha \beta \gamma \delta \f} &=& 12 e^{Z+3W} J_{[\alpha \beta} J_{\gamma \delta]} \\
F_{(5)abc\alpha \beta} &=& \frac{1}{2} 
e^{W+2U} \epsilon_{abc}^{\;\;\;\;\;de} \left(K_{2;de} J_{\alpha \beta} + L_{2;de} \Omega_{\alpha \beta} + L^*_{2;de} \bar{\Omega}_{\alpha \beta} \right)   \\
F_{(5)ab\alpha \beta \f} &=&  e^{W+2U}  \left(K_{2;ab} J_{\alpha \beta} + L_{2;ab} \Omega_{\alpha \beta} + L^*_{2;ab} \bar{\Omega}_{\alpha \beta} \right).
\end{eqnarray} 

\noindent Similarly for the components of $F_{(3)}$ with respect to the ten-dimensional frame we find
\begin{align}
\label{F3 components 1}F_{(3)abc} &= e^{-3W}G_{3\,abc}\\
F_{(3)ab\f} &= e^{-2W - V}G_{2\, ab}\\
F_{(3)a\alpha\beta} &= e^{-W-2U}\bigl[G_{1\, a}J_{\alpha\beta} + \left(N_{1\, a}\Omega_{\alpha\beta}+\mbox{ c.c.}\right)\bigr]\\
\label{F3 components 4}
F_{(3)\alpha\beta \f}&= e^{-2U-V}\bigl[G_{0}J_{\alpha\beta} + \left(N_{0}\Omega_{\alpha\beta} +\mbox{ c.c.}\right)\bigr],
\end{align}

\noindent with an analogous expression for $H_{(3)}$.

\subsection{Clifford algebra}\label{Appendix:Clifford}
 We choose the following basis for the $D=10$ Clifford algebra:
\beqn
\Gamma^a&=&\gamma^a\otimes \mathds{1}_4\otimes \sigma_1\\
\Gamma^\alpha &=& \mathds{1}_4\otimes \gamma^\alpha\otimes \sigma_2\\
\Gamma_\f&=& \mathds{1}_4\otimes \gamma_\f \otimes \sigma_2\, ,
\eeqn

\noindent where $a=0,1,...,4$, $\alpha=1,...,4$, whence\footnote{We take $\gamma^4=i\gamma^{0123}$ in $C\ell(4,1)$. There is of course the opposite sign choice, leading to an inequivalent irrep of $C\ell(4,1)$.} 
\beqn
\Gamma^{ab}&=& \gamma^{ab}\otimes \mathds{1}_4\otimes \mathds{1}_2\\
\Gamma^{\alpha\beta}&=& \mathds{1}_4\otimes \gamma^{\alpha\beta}\otimes \mathds{1}_2\\
\Gamma_{11}&=& -\Gamma^0\Gamma^1...\Gamma^9=\mathds{1}_4\otimes \mathds{1}_4\otimes \sigma_3\, . \label{defgamma11}
\eeqn

\noindent The $\gamma^a$ generate $C\ell(4,1)$ while the $\gamma^\alpha$ generate $C\ell(4,0)$. We have $\gamma^{01234}=-i \mathds{1}_4$ in $C\ell(4,1)$ and  $\gamma_\f =-\gamma^1\gamma^2\gamma^3\gamma^4$ in $C\ell(4,0)$.

Notice that $\gamma^{abcde} = i \epsilon^{abcde}_5$. Some useful identities involving the $C\ell(4,1)$ gamma matrices are then
\begin{align}
\epsilon_{abcde} \gamma^{abcde} &= -i5!\, ,&  \epsilon^e_{\;\;abcd} \gamma^{abcd} &= -i4! \gamma^e\, ,\\
\epsilon^{de}_{\;\;\;\;abc} \gamma^{abc} &= +i3! \gamma^{de}\, ,& \epsilon^{cde}_{\;\;~\;\;ab} \gamma^{ab} &= +i2! \gamma^{cde}\, .
\end{align}

\noindent It is also useful to notice that the K\"ahler form on $KE$ satisfies
\beqn
J_{\alpha\beta}J_{\gamma\delta}\epsilon^{\alpha\beta\gamma\delta}=8\, , \qquad J_{\alpha\beta}J_{\gamma\delta}\gamma^{\alpha\beta\gamma\delta}=-8\gamma_\f\, , \qquad 
J_{\alpha\beta}J_{\gamma\delta}\gamma^{\beta\gamma\delta}= -2\gamma_\alpha\gamma_\f\, .
\eeqn

\subsection{Charge conjugation conventions}\label{Appendix:ChargeConjugation}
In $d=5$ dimensions with signature $(-,+,+,+,+)$ we can define unitary intertwiners $B_{4,1}$ and $C_{4,1}$ (the charge conjugation matrix), unique up to a phase, satisfying
\begin{align}
B_{4,1}\gamma^aB_{4,1}^{-1}={-\gamma^a}^*\, ,\qquad B_{4,1}^{T} = -B_{4,1}\, ,\qquad
B_{4,1}^{*}B_{4,1} =-\mathds{1}\, , 
\end{align}

\noindent and 
\begin{align}\label{summary intertwiner C summ}
    C_{4,1}\gamma_{a}C_{4,1}^{-1} = \gamma^{T}_{a}\, ,\qquad  C_{4,1}^{T} = -C_{4,1}\, ,\qquad
    C_{4,1} = B_{4,1}^{T}\gamma_{0}=-B_{4,1}\gamma_{0}\, .  
\end{align}

\noindent If $\psi$ is any spinor in $(4+1)$ dimensions, its charge conjugate $\psi^{\conj}$  is then defined  as
\begin{equation}\label{4 + 1 cc}
\psi^{\conj} = B_{4,1}^{-1}\psi^{*}=B_{4,1}^{\dagger}\psi^{*} = -\gamma_{0}C_{4,1}^{\dagger}\psi^{*}\, .
\end{equation}

\noindent In (4+1) dimensions it is not possible to define Majorana spinors satisfying $\psi^{\conj} = \psi$. It is possible, however, to define \textit{symplectic Majorana} spinors. These satisfy $\psi^{\conj}_{i} = \Omega_{ij}\psi_{j}$, where $\Omega_{ij}$ is the USp(4)-invariant symplectic form. This fact becomes particulary relevant when dealing with $N=4$ supergravity in $d=5$ dimensions, inasmuch as the symplectic Majorana spinors allow to make the action of the $R$-symmetry manifest.

In analogy with  \eqref{4 + 1 cc}, we can define the charge conjugates of a spinor $\Psi$ in (9+1) dimensions and a spinor $\varepsilon$ in $5$ Euclidean dimensions as
\begin{alignat}{7}
\Psi^{\conj} &= B_{9,1}^{-1}\Psi^{*}\, ,&  &\phantom{3445}& &\mbox{where}& &\phantom{3445}& B_{9,1}\Gamma_{M} B_{9,1}^{-1} &=\Gamma_{M}^{*}\, , &\phantom{3445} B_{9,1}^{T} &= B_{9,1}  
\\
\varepsilon^{\conj} &= B_{5}^{-1}\varepsilon^{*}\, ,&  &\phantom{3445}& &\mbox{where}& &\phantom{3445}& B_{5}\gamma_{\alpha} B_{5}^{-1} &=\gamma_{\alpha}^{*}\, ,&\phantom{3445} B_{5}^{T} &= -B_{5}\, ,
\end{alignat}

\noindent where $B_{5}$ and $B_{9,1}$ are the corresponding unitary intertwiners. We then find
\begin{equation}
B_{9,1} = B_{4,1}\otimes B_{5}\otimes \sigma_{3}\, .
\end{equation}

\noindent Notice that $B_{5}$ is unitary and antisymmetric, and therefore for a spinor $\varepsilon$ in five Euclidean dimensions we have $\left(\varepsilon^\conj\right)^\conj = -\varepsilon$. In particular, in terms of the gauge-covariantly constant spinors $\varepsilon_{\pm}$ introduced in section \ref{section:Ansatz}, we have that defining $\varepsilon_{-}$ as the charge conjugate of $\varepsilon_{+}$, this is $e^{-\frac{3i}{ 2}\chi}\varepsilon_{-}\equiv\left(e^{\frac{3i}{2}\chi }\varepsilon_{+}\right)^{\conj}$, implies that $\left(e^{-\frac{3i}{2}\chi }\varepsilon_{-}\right)^{\conj}= -e^{\frac{3i}{2}\chi }\varepsilon_{+}$. We also define the unitary intertwiner $C_{9,1}$ (the charge-conjugation matrix) in $(9 +1)$ dimensions, which satisfies
\begin{align}
 C_{9,1}\Gamma_{M}C_{9,1}^{-1} = -\Gamma_{M}^{T}\, \qquad\qquad C_{9,1} = B_{9,1}^{T}\Gamma_{0} = B_{9,1}\Gamma_{0}\, .
\end{align}

\noindent Notice that defining $\Psi^{\conj}$ in the (9+1)-dimensional space by using the intertwiner $B_{9,1}$ introduced above (as opposed to using an intertwiner $B_{9,1}^{-}$ satisfying $B_{9,1}^{-}\Gamma_{M}B_{9,1}^{-\dagger} =  -\Gamma^{*}_{M}$) allows one to choose a basis, if so desired, where the charge conjugation operation in $D=10$ reduces to complex conjugation. In this basis all the $C\ell(9,1)$ gamma-matrices are real, with $B_{9,1} = \mathds{1}$ and a corresponding (9+1) charge-conjugation matrix $C_{9,1} = B_{9,1}^{T}\Gamma_{0}=\Gamma_{0}$.

\section{Type IIB supergravity}\label{Appendix:typeIIB}
In this appendix we briefly review the field content and equations of motion of type IIB supergravity \cite{Schwarz:1983qr,Howe:1983sra}. We follow the conventions of \cite{Gauntlett:2010vu}, \cite{Gauntlett:2005ww}, \cite{Gauntlett:2007ma} closely, and adapt our fermionic conventions accordingly. 
 
\subsection{Bosonic content and equations of motion}\label{Appendix:typeIIBbosonic}
In the $SU(1,1)$ language of \cite{Schwarz:1983qr}, the bosonic content of type IIB supergravity includes the metric, a complex scalar $B$, ``composite" complex $1$-forms $P$ and $Q$ (that can be written in terms of $B$), a complex $3$-form $G$, and a real self dual five-form $F_{(5)}$. The corresponding equations of motion read (to linear order in the fermions)
\beqn
D*P &=& -\frac14G\wedge *G\label{eom SU(1,1) first}\\
D*G &=& P\wedge *G^*-iG\wedge F_{(5)}\\
 R_{MN} &=& P_M P_N^*+P_NP^*_M
      + \frac{1}{96}F_{(5)MP_1P_2P_3P_4}F_{(5)N}^{\phantom{(5)N}P_1P_2P_3P_4}  \nonumber \\ &&
       + \frac{1}{8}\left(
         G_M{}^{P_1P_2}G^*_{NP_1P_2} + G_N{}^{P_1P_2}G^*_{MP_1P_2}
         - \frac{1}{6}g_{MN}G^{P_1P_2P_3}G^*_{P_1P_2P_3} \right)\label{eom SU(1,1) last}
\eeqn

\noindent together with the self-duality condition $*F_{(5)} =F_{(5)}$. Similarly, the Bianchi identities read
\begin{align}
 \label{IIBBianchiF}  dF_{(5)}-\frac{i}{2} G \wedge G^* &=0 \\
 \label{IIBBianchiG} DG+P \wedge G^* &=0 \\
 \label{IIBBianchiP} DP &=0\, .
\end{align}

\noindent In this language there is a manifest local $U(1)$ invariance and $Q$ is the corresponding gauge field, with field-strength $d Q = -iP\wedge P^*$. Similarly, $G$ has charge 1 and $P$ has charge 2 under the $U(1)$, so $D*G \equiv d* G -i Q\wedge *G$ and $D*P \equiv d*P-2iQ \wedge *P$. Notice that Einstein's equation \eqref{eom SU(1,1) last} has been rewritten by using the trace condition $R = 2P^{R}P^{*}_{R} + \frac{1}{24}G^{P_1P_2P_3}G^{*}_{P_1P_2P_3}\,$.

In the body of the paper we have worked in the $SL(2,\mathds{R})$ language which is more familiar to string theorists. The translation between the two formalisms involves a gauge-transformation and field-redefinitions.\footnote{The gauge transformation has the form $P \to e^{2i\theta}P$, $Q \to Q + d\theta$, $G \to e^{\frac{i}{2}\theta}G$, where $\theta$ is a $\tau$-dependent phase. These phases are then absorbed by a redefinition of the fermions. More details can be found in \cite{DeWolfe:2002nn,Chiodaroli:2009yw}, for example.} Here we just quote the result that links this formalism with the fields used in the rest of the paper. Writing the axion-dilaton $\tau$ and the NSNS and RR $3$-forms $H_{(3)}$ and $F_{(3)}$ as 
\begin{equation}\label{definition axion dilaton and 3-forms}
\tau \equiv C_{(0)} + ie^{-\Phi}\, ,\qquad F_{(3)} = dC_{(2)}-C_{(0)}dB_{(2)}\, ,\qquad H_{(3)} = dB_{(2)}\,,
\end{equation}

\noindent for the 3-form $G$ we have\footnote{Note that our forms $F_{(3)}$ and $G$ are related to the traditional string theory forms $F_{(3)st} = dC_{(2)}$ and $G_{st} = F_{(3)st} - \tau H_{(3)}$ by $F_{(3)} = F_{(3)st} - C_{(0)}H_{(3)}$ and $G = -iG_{st}/\sqrt{\mbox{Im}\tau}$ . It's not our fault.}  \cite{Gauntlett:2005ww}
\begin{equation}\label{definition G}
G = ie^{\Phi/2}\left(\tau dB - dC_{(2)}\right) =  -\left(e^{-\Phi/2}H_{(3)} + ie^{\Phi/2}F_{(3)}\right),
\end{equation}

\noindent and similarly
\begin{equation}
P = \frac{i}{2}e^{\Phi}d\tau\, ,\qquad Q = -\frac{1}{2}e^{\Phi}dC_{(0)}\, .
\end{equation}

\noindent In terms of these fields, the equations of motion \eqref{eom SU(1,1) first}-\eqref{eom SU(1,1) last} become \cite{Gauntlett:2010vu} (to linear order in the fermions)
\begin{eqnarray}
0&=& d(e^\Phi *F_{(3)})  - F_{(5)} \wedge H_{(3)}\label{eomF3} \\[8pt]
0&=& d(e^{2\Phi} *F_{(1)})  +e^{\Phi} H_{(3)} \wedge *F_{(3)}  \label{eomF1} \\[8pt]
0&=& d(e^{-\Phi} *H_{(3)})  -e^{\Phi} F_{(1)} \wedge *F_{(3)} -F_{(3)} \wedge F_{(5)}   \label{eomH3} \\[8pt]
0&=& d*d\Phi -e^{2\Phi} F_{(1)} \wedge *F_{(1)} +\frac{1}{2} e^{-\Phi} H_{(3)} \wedge *H_{(3)} -\frac{1}{2} e^{\Phi} F_{(3)} \wedge *F_{(3)}  \label{eomPhi}\\[8pt]
 R_{MN} &=& \frac{1}{2} e^{2\Phi} \nabla_M C_{(0)} \nabla_N C_{(0)} + \frac{1}{2} \nabla_M \Phi \nabla_N \Phi
      + \frac{1}{96}F_{MP_1P_2P_3P_4}F_{N}^{\phantom{N}P_1P_2P_3P_4}  \nonumber \\  
     && + \frac{1}{4}e^{-\Phi}\left(
         H_M{}^{P_1P_2} H_{NP_1P_2}
         - \frac{1}{12}g_{MN}H^{P_1P_2P_3}H_{P_1P_2P_3} \right) \nonumber \\ && 
      + \frac{1}{4}e^{\Phi}\left(
         F_M{}^{P_1P_2} F_{NP_1P_2}
         - \frac{1}{12}g_{MN}F^{P_1P_2P_3}F_{P_1P_2P_3} \right)  \label{IIBEinstein}
\end{eqnarray}

\noindent while the Bianchi identities \eqref{IIBBianchiF}-\eqref{IIBBianchiP} now read
\begin{eqnarray}
&& dF_{(5)} + F_{(3)} \wedge  H_{(3)} =0 \label{F5} \\
&& dF_{(3)}  + F_{(1)} \wedge H_{(3)}=0 \label{F3} \\
&& dF_{(1)} =0 \label{F1} \\
&& dH_{(3)} =0\, . \label{H3}
\end{eqnarray}
\noindent These identities are solved by writing $F_{(5)} = dC_{(4)} -C_{(2)} \wedge  H_{(3)}$,  $F_{(1)} = dC_{(0)}$, together with $H_{(3)} = dB_{(2)}$ and $F_{(3)} = dC_{(2)} -C_{(0)}  dB_{(2)}$ as in \eqref{definition axion dilaton and 3-forms}.

\subsection{Fermionic content and equations of motion}
Our conventions for the type IIB fermionic sector are based on those of \cite{DeWolfe:2002nn}, \cite{Argurio:2006my}, with slight modifications needed to conform with our bosonic conventions. The type IIB fermionic content consists of a chiral dilatino $\lambda$ and a chiral gravitino $\Psi$, with equations of motion given by (to linear order in the fermions) 
\beqn
\label{10d dilatino eom}\hat\Dsl\lambda &=& \frac{i}{8}\Fsl_{(5)}\lambda + {\cal O}(\Psi^{2})\\
\label{10d gravitino eom}\Gamma^{ABC}\hat D_B\Psi_C
&=&-\frac{1}{8}\Gsl^*\Gamma^A\lambda+\fudge \frac{1}{2}\Psl\Gamma^A\lambda^\conj + {\cal O}(\Psi^{3})
\eeqn

\noindent Here, $\hat{D}$ denotes the flux-dependent supercovariant derivative, which acts as follows:
\begin{eqnarray}
\label{DilSuper}
\hat{\Dsl}\lambda &=& \left(\hat{\slashed{\nabla}} -\frac{3i}{2}\slashed{Q}\right)\lambda -\frac{1}{4}\Gamma^A\Gsl\Psi_A-\Gamma^A\Psl\Psi^\conj_A  \\ 
\hat D_B\Psi_C &=& \left(\hat{\nabla}_B -\frac{i}{2}Q_B\right)\Psi_C+\frac{i}{16}\Fsl_{(5)}\Gamma_B\Psi_C-\frac{1}{16}S_{B}\Psi^\conj_C \, ,
\end{eqnarray}

\noindent where $\hat{\nabla}_{B}$ denotes the ordinary 10-$d$ spinor covariant derivative and we have defined 
\begin{eqnarray}
\label{GravSuper}
S_B \equiv\frac{1}{6}\left({\Gamma_B}^{DEF}G_{DEF}-9\Gamma^{DE}G_{BDE}\right).
\end{eqnarray}

The gravitino and dilatino have opposite chirality in $d=10$, and we choose $\Gamma_{11}\Psi_A=-\Psi_A$, $\Gamma_{11}\lambda=+\lambda$. Since $F_{(5)}$ is self-dual, our conventions then imply $\Fsl_{(5)}=-\Gamma_{11}\Fsl_{(5)}$. Thus, for any spinor $\varepsilon$ satisfying $\Gamma_{11}\varepsilon=-\varepsilon$ we have $\Fsl_{(5)}\varepsilon=0$ and $\Fsl_{(5)}\Gamma_A\varepsilon=\{\Fsl_{(5)},\Gamma_A\}\varepsilon=\frac{1}{12}F_{(5)ACDEF}\Gamma^{CDEF}\varepsilon$.
The corresponding SUSY variations of the fermions read
\beqn
\delta\lambda &=& \Psl\varepsilon^\conj+\frac{1}{4}\Gsl\varepsilon\\
\delta\Psi_A&=& \left(\hat{\nabla}_A -\frac{i}{2}Q_A\right)\varepsilon+\frac{i}{16}\Fsl_{(5)}\Gamma_A\varepsilon-\frac{1}{16}S_{A}\varepsilon^\conj\, .
\eeqn

\section{$d=5$ Equations of motion}\label{Appendix:EOM}
In this appendix we present the dimensional reduction of the fermionic equations of motion in full detail, and rewrite them in final form in terms of the fields  \eqref{canonical fields 1}-\eqref{canonical fields 4} which possess diagonal kinetic terms in the effective action. In the calculations below we encounter a number of expressions involving $\varepsilon_{\pm}$ that need evaluation. We collect them here:
\begin{align}
\Jsl\varepsilon_+&=\frac12 i Q\varepsilon_+=2i\varepsilon_+
&
\Jsl\varepsilon_-&=\frac12 i Q\varepsilon_-=-2i\varepsilon_-
\\
\Omsl\varepsilon_-e^{-\frac32i\chi}&=4\varepsilon_+e^{\frac32i\chi}
&
\Ombsl\varepsilon_+e^{\frac32i\chi}&=-4\varepsilon_-e^{-\frac32i\chi}
\\
\gamma^\alpha\gamma_\alpha\varepsilon_+&=4\varepsilon_+
&\label{gammagammaeps}
\gamma^\alpha\Jsl\gamma_\alpha\varepsilon_+&=\gamma^{\bar\alpha}\Omsl\gamma_{\bar\alpha}\varepsilon_-=0\, .
\end{align}

\subsection{Reduction of the dilatino equation of motion}
We begin by performing the reduction of the $D=10$ equation of motion for the dilatino, as given in \eqref{10d dilatino eom}.
\subsubsection{Derivative operator}
We first reduce the 10-$d$ derivative operator $\hat{\nabla}_{A} - (3i/2)Q_{A}$ acting on the dilatino. Defining
\begin{equation}
e^{W}\left(\hat{\nablasl} - \frac{3i}{2}\Qsl\right)\lambda \equiv \mathcal{L}_{\lambda}^{+}\otimes \varepsilon_{+}e^{\frac{3i}{2}\chi}\otimes  u_{-} +  \mathcal{L}_{\lambda}^{-}\otimes \varepsilon_{-}e^{-\frac{3i}{2}\chi}\otimes  u_{-}
\end{equation}

\noindent we find
\begin{align}
\mathcal{L}_{\lambda}^{\pm} ={}& \left(
\Dsl+\frac{1}{2}\pasl W+\frac34 ie^{\phi} (\pasl a)\right)\lambda^{(\pm)}
+\frac14i\Sigma^{-2}\Fsl_{2}\lambda^{(\pm)}
\mp \left(e^{-4U}\Sigma^{-1}+\frac32\Sigma^2\right)\lambda^{(\pm)}\, ,
\end{align}
\noindent where $\Dsl \lambda^{(\pm)}=\left(\nablasl \mp\frac{3}{2}i\Asl_1\right)\lambda^{(\pm)}$ is the gauge-covariant five-dimensional connection acting on $\lambda^{(\pm)}$.

\subsubsection{Couplings}
We now reduce the various terms involving the couplings of the dilatino, including the flux-dependent terms in the supercovariant derivative. Defining
\begin{equation}
e^{W}\left(\frac{i}{8}\Fsl_{(5)}\lambda +\frac{1}{4}\Gamma^A\Gsl\Psi_A\right) \equiv \mathcal{R}_{1\lambda}^{+}\otimes \varepsilon_{+}e^{\frac{3i}{2}\chi}\otimes  u_{-} +  \mathcal{R}_{1\lambda}^{-}\otimes \varepsilon_{-}e^{-\frac{3i}{2}\chi}\otimes  u_{-}
\end{equation}

\noindent we find 
\begin{align}
\mathcal{R}_{1\lambda}^{(\pm)}
={}&
 e^{Z+4W} \lambda^{(\pm)} 
 - \frac12i e^{-4U} \slashed{K}_1\lambda^{(\pm)}\mp\frac12i\Sigma\slashed{K}_2\lambda^{(\pm)} 
\mp \Sigma \slashed{L}_2^{(\pm)} \lambda^{(\mp)}
\nonumber\\
&
-\frac14i\gamma^{a}\cgsl{3}\psi^{(\pm)}_a
-\frac14  \gamma^a\cgsl{2}\psi^{(\pm)}_a
+\frac14\cgsl{3}\left(\varphi^{(\pm)}+4\rho^{(\pm)}\right)
-\frac14i \cgsl{2}\left(\varphi^{(\pm)}-4\rho^{(\pm)}\right)
\nonumber\\
& 
\pm\frac12 \gamma^a\cgsl{1}\psi^{(\pm)}_a
\pm\frac12i\cgsl{1}\varphi^{(\pm)}  
\mp i\gamma^a\cnpmsl{1}\psi^{(\mp)}_a
\pm\cnpmsl{1}\varphi^{(\mp)} 
\nonumber\\
& 
\mp \gamma^a\cnpm{0}{}{}\psi^{(\mp)}_a
\mp i\cnpm{0}{}{}\varphi^{(\mp)}\, ,
\end{align}

\noindent where we have introduced the notation $\slashed{L}_{2}^{(+)} = (1/2!)L_{2\, ab}\gamma^{ab}$ and $\slashed{L}_{2}^{(-)} = (1/2!)L^{*}_{2\, ab}\gamma^{ab}$. Similarly, defining
\begin{equation}
e^{W}\Gamma^A\Psl\Psi^\conj_A \equiv \mathcal{R}_{2\lambda}^{+}\otimes \varepsilon_{+}e^{\frac{3i}{2}\chi}\otimes  u_{-} +  \mathcal{R}_{2\lambda}^{-}\otimes \varepsilon_{-}e^{-\frac{3i}{2}\chi}\otimes  u_{-}
\end{equation}

\noindent we obtain
\begin{align}
\mathcal{R}_{2\lambda}^{(\pm)}
=&{}
 \pm\Psl\psi^{(\mp)\conj}_a \pm i\Psl\left(4\rho^{(\mp)\conj}+\varphi^{(\mp)\conj}\right).
\end{align}

\noindent where, in a slight abuse of notation, $\Psl = (1/2)\left(\pasl \phi + ie^{\phi}\pasl a\right)$ when appearing in 5-$d$ equations. In terms of the quantities computed above, the 10-$d$ dilatino equation reduces to two equations for the five-dimensional fields, given by
\begin{equation}
\mathcal{L}_{\lambda}^{(\pm)}-\mathcal{R}_{1\lambda}^{(\pm)}-\mathcal{R}_{2\lambda}^{(\pm)}=0\, .
\end{equation}

\subsection{Reduction of the gravitino equation of motion}
We now reduce the equation of motion for the $D=10$ gravitino, as given in \eqref{10d gravitino eom}.
\subsubsection{Derivative operator}
Here we define
\begin{align}
e^{W}\Gamma^{aBC}\left(\hat{\nabla}_B -\frac{i}{2}Q_B\right)\Psi_{C} &= \mathcal{L}^{(+)a} \otimes \varepsilon_{+}e^{\frac{3i}{2}\chi}\otimes  u_{+}  + \mathcal{L}^{(-)a} \otimes \varepsilon_{-}e^{-\frac{3i}{2}\chi}\otimes  u_{+} 
\\
e^{W}\tilde{\sigma}_{2}\Gamma_{\alpha}\Gamma^{\alpha BC}\left(\hat{\nabla}_B -\frac{i}{2}Q_B\right)\Psi_{C} &=\mathcal{L}^{(+)}_{base} \otimes \varepsilon_{+}e^{\frac{3i}{2}\chi}\otimes  u_{+} +\mathcal{L}^{(-)}_{base} \otimes \varepsilon_{-}e^{-\frac{3i}{2}\chi}\otimes  u_{+} 
\\
 e^{W}\Gamma^{\f BC}\left(\hat{\nabla}_B -\frac{i}{2}Q_B\right)\Psi_{C} &=\mathcal{L}_{\f}^{(+)} \otimes \varepsilon_{+}e^{\frac{3i}{2}\chi}\otimes  u_{+} + \mathcal{L}_{\f}^{(-)} \otimes \varepsilon_{-}e^{-\frac{3i}{2}\chi}\otimes  u_{+}
\end{align}
\noindent where $\tilde\sigma_{2}\equiv \mathds{1}_{4}\otimes \mathds{1}_{4}\otimes\sigma_2$. Then, for the components of the derivative operator in the external manifold directions we find
\begin{align}
\mathcal{L}^{(\pm)a}
={}&
\gamma^{abc}\left(D_{b}+\frac12\pa_bW+\frac14ie^{\phi} (\pa_b a) \right)\psi_{c}^{(\pm)}
\nonumber\\
&
-\frac14i\Sigma^{-2}\gamma^{[c}\Fsl_{2}\gamma^{a]}\psi^{(\pm)}_{c}  \mp\left( \Sigma^{-1}e^{-4U}
+\frac{3}{2}\Sigma^{2}\right)\gamma^{ab}\psi_b^{(\pm)}
\nonumber\\
&
-4i\gamma^{ab}\left[D_b+\frac12\pa_b W+\frac14ie^{\phi} (\pa_b a)\right]\rho^{(\pm)}
-i(\Sigma^{-1}\pasl\Sigma)\gamma^{a}\rho^{(\pm)}
+4i(\pasl U)\gamma^{a}\rho^{(\pm)}
\nonumber\\
&
\pm 2i\left(3\Sigma^{2} +\Sigma^{-1}e^{-4U}\right)\gamma^a\rho^{(\pm)} -\frac{1}{2}\Sigma^{-2}F_{2\, bd}\gamma^{b}\gamma^{a}\gamma^{d}\rho^{(\pm)}
\nonumber\\
&
-i\gamma^{ab}\left[D_b+\frac12\pa_b W+\frac14ie^{\phi} (\pa_b a)\right]\varphi^{(\pm)}
-i(\Sigma^{-1}\pasl\Sigma)\gamma^{a}\varphi^{(\pm)}
\nonumber\\
&
\pm 2i\Sigma^{-1}e^{-4U}\gamma^a\varphi^{(\pm)}+\frac14\Sigma^{-2}F_{2\, bc}\gamma^c\gamma^{ab}\varphi^{(\pm)}\, .
\end{align}

\noindent Similarly, the components in the direction of the KE base yield
\begin{align}
\mathcal{L}_{base}^{(\pm)}
={}&
-4i\gamma^{ab}\left[D_{a}+\frac12\pa_a W+\frac14ie^{\phi} (\pa_a a)\right]\psi^{(\pm)}_{b}
+i\gamma^{b}(\Sigma^{-1}\pasl\Sigma)\psi^{(\pm)}_b
-4i\gamma^{b}(\pasl U)\psi^{(\pm)}_b
\nonumber\\
&
+\frac12\Sigma^{-2}F_{2\, da}\gamma^{a}\gamma^{b} \gamma^d\psi_{b}^{(\pm)}
\pm 2i\left( \Sigma^{-1}e^{-4U}+ 3\Sigma^{2}\right)\gamma^{b}\psi_b^{(\pm)}
\nonumber\\
&
-12\left[\Dsl +\frac{1}{2}(\pasl W) +\frac14ie^{\phi} (\pasl a)\right]\rho^{(\pm)} \pm 2\left(2\Sigma^{-1}e^{-4U} + 9\Sigma^{2}\right)\rho^{(\pm)} -3i\Sigma^{-2}\Fsl_{2}\rho^{(\pm)}
\nonumber\\
&
-4 \left[\Dsl+\frac12\pasl W+\frac14ie^{\phi} (\pasl a)-\frac34(\Sigma^{-1}\pasl\Sigma)-(\pasl U)\right]\varphi^{(\pm)}
\nonumber\\
&
\pm 2\Sigma^{-1}e^{-4U}\varphi^{(\pm)} - 2i\Sigma^{-2}\Fsl_{2}\varphi^{(\pm)}\, .
\end{align}

\noindent Finally, for the fiber component of the derivative operator we obtain
\begin{align}
\mathcal{L}_{\f}^{(\pm)}
=&{}
-i\gamma^{ab}\left[D_{a}+\frac12\pa_aW+\frac14ie^{\phi} (\pa_a a)\right]\psi^{(\pm)}_b
+i\gamma^b(\Sigma^{-1}\pasl\Sigma)\psi^{(\pm)}_b
\nonumber\\
&
\pm 2i\Sigma^{-1}e^{-4U}\gamma^{b}\psi_{b}^{(\pm)}+\frac14\Sigma^{-2}F_{2\, da} \gamma^{ab}\gamma^d\psi^{(\pm)}_b
\nonumber\\
&
-4 \left[\Dsl+\frac12\pasl W +\frac14ie^{\phi} (\pasl a)+\frac34(\Sigma^{-1}\pasl\Sigma)  +\pasl U\right]\rho^{(\pm)}
\nonumber\\
&
\pm 2\Sigma^{-1}e^{-4U}\left( 2\varphi^{(\pm)}+\rho^{(\pm)}\right) 
-i\Fsl_{2}\Sigma^{-2}\left(\varphi^{(\pm)} +2\rho^{(\pm)}\right).
\end{align}

\subsubsection{Couplings}
Next, define
\begin{align}
e^{W}\left(-\frac{1}{8}\Gsl^{*} \Gamma^{a}\lambda -\frac{i}{16}\Gamma^{aBC}\Fsl_{(5)}\Gamma_{B}\Psi_{C}\right) &= \mathcal{R}_{1}^{(+)a} \otimes \varepsilon_{+}e^{\frac{3i}{2}\chi}\otimes  u_{+}\nonumber\\
&\,   + \mathcal{R}_{1}^{(-)a} \otimes \varepsilon_{-}e^{-\frac{3i}{2}\chi}\otimes  u_{+} 
\\
e^{W}\left(-\frac{1}{8}\tilde{\sigma}_{2}\Gamma_{\alpha}\Gsl^{*} \Gamma^{\alpha}\lambda -\frac{i}{16}\tilde{\sigma}_{2}\Gamma_{\alpha}\Gamma^{\alpha BC}\Fsl_{(5)}\Gamma_{B}\Psi_{C}\right) &=\mathcal{R}^{(+)}_{1\, base} \otimes \varepsilon_{+}e^{\frac{3i}{2}\chi}\otimes  u_{+}\nonumber\\
&\, +\mathcal{R}^{(-)}_{1\,base} \otimes \varepsilon_{-}e^{-\frac{3i}{2}\chi}\otimes  u_{+} 
\\
e^{W}\left(-\frac{1}{8}\Gsl^{*} \Gamma^{\f}\lambda -\frac{i}{16}\Gamma^{\f BC}\Fsl_{(5)}\Gamma_{B}\Psi_{C}\right)&=\mathcal{R}_{1\,\f}^{(+)} \otimes \varepsilon_{+}e^{\frac{3i}{2}\chi}\otimes  u_{+} \nonumber\\
&\, + \mathcal{R}_{1\, \f}^{(-)} \otimes \varepsilon_{-}e^{-\frac{3i}{2}\chi}\otimes  u_{+}\, .
\end{align}

\noindent We find
\begin{align}
{\cal R}_1^{(\pm)a}
={}& \left(-\frac18i\cgtsl{3} \pm\frac14 \cgtsl{1} -\frac18 \cgtsl{2} \right)\gamma^a\lambda^{(\pm)}
 \mp\left(\frac12i \cntpmsl{1}+\frac12 \cntpm{0}{}{}\right)\gamma^a\lambda^{(\mp)}\nonumber\\
&
+ e^{Z+4W} \gamma^{ba} \psi^{(\pm)}_b
- \frac12i e^{-4U} \gamma^{[b}\slashed K_1\gamma^{a]}\psi^{(\pm)}_b
+  e^{-4U}\{ \slashed{K}_{1},\gamma^{a}\}\rho^{(\pm)}
- \frac14 e^{-4U}[ \slashed{K}_{1},\gamma^a]\varphi^{(\pm)}
\nonumber\\
&
 \mp\frac12i\Sigma\gamma^{[b}\slashed{K}_2\gamma^{a]}\psi^{(\pm)}_b
 \mp \Sigma\gamma^{[b}\slashed{L}^{(\pm)}_2\gamma^{a]} \psi^{(\mp)}_b
-\frac14\Sigma \left(\pm[\slashed{K}_{2},\gamma^a]\varphi^{(\pm)}\mp 2i [\slashed{L}_{2}^{(\pm)},\gamma^a]\varphi^{(\mp)}\right)
\nonumber\\
&
+\Sigma \gamma^a\left(\pm\slashed{K}_2\rho^{(\pm)}  \mp 2i\slashed{L}_2^{(\pm)} \rho^{(\mp)} \right)
\\
\nonumber
\\
\mathcal{R}^{(\pm)}_{1\, base}
={}&
\left(\frac12 \cgtsl{3}+\frac{i}{2}\cgtsl{2}\right)\lambda^{(\pm)}+e^{-4U}\{\gamma^b, \slashed{K}_1\}\psi^{(\pm)}_b
-6ie^{-4U} \slashed{K}_1\rho^{(\pm)}
 +4 e^{Z+4W}( \varphi^{(\pm)}+3\rho^{(\pm)})
\nonumber\\
&
-\Sigma  \left[\pm i\slashed{K}_{2}\left(i\gamma^{a}  \psi^{(\pm)}_a+\varphi^{(\pm)}+2\rho^{(\pm)}\right) \pm 2 \slashed{L}^{(\pm)}_{2}\left(i \gamma^{a} \psi^{(\mp)}_a +\varphi^{(\mp)}+2\rho^{(\mp)}\right)\right] 
 \\
 \nonumber
 \\
{\cal R}_{1\, \f}^{(\pm)}
={}& 
\left(\frac18\cgtsl{3} \pm \frac14i\cgtsl{1}-\frac18i  \cgtsl{2}\right)\lambda^{(\pm)} 
 \pm \left(\frac12\cntpmsl{1}-\frac12i \cntpm{0}{}{}\right)\lambda^{(\mp)}
\nonumber\\
&-\frac14 e^{-4U}[ \gamma^b,\slashed{K}_1]\psi^{(\pm)}_b
\mp\frac14\Sigma[\gamma^b, \slashed{K}_2]\psi^{(\pm)}_b
\pm  \frac12i\Sigma[\gamma^b, \slashed{L}^{(\pm)}_2]\psi^{(\mp)}_b
\nonumber\\
&
 +4 e^{Z+4W} \rho^{(\pm)}   \mp i\Sigma\slashed{K}_2\rho^{(\pm)}  \mp 2\Sigma\slashed{L}^{(\pm)}_2 \rho^{(\mp)}\, .
\end{align}

We now reduce the couplings to the charge conjugate spinors in the gravitino equation. We write
\begin{align}
\frac{1}{2}e^{W}\Psl \Gamma^{a}\lambda^{\conj}+\frac{1}{16}e^{W}\Gamma^{aBC}S_{B}\Psi^{\conj}_{C} &= \mathcal{R}_{2}^{(+)a} \otimes \varepsilon_{+}e^{\frac{3i}{2}\chi}\otimes  u_{+}
\nonumber\\
&\,  + \mathcal{R}_{2}^{(-)a} \otimes \varepsilon_{-}e^{-\frac{3i}{2}\chi}\otimes  u_{+} 
\\
\frac{1}{2}e^{W}\tilde{\sigma}_{2}\Gamma_{\alpha}\Psl \Gamma^{\alpha}\lambda^{\conj} +\frac{1}{16}e^{W}\tilde{\sigma}_{2}\Gamma_{\alpha}\Gamma^{\alpha BC}S_{B}\Psi^{\conj}_{C} &=\mathcal{R}^{(+)}_{2\, base} \otimes \varepsilon_{+}e^{\frac{3i}{2}\chi}\otimes  u_{+}
\nonumber\\
&
\, +\mathcal{R}^{(-)}_{2\,base} \otimes \varepsilon_{-}e^{-\frac{3i}{2}\chi}\otimes  u_{+} 
\\
\frac{1}{2} e^{W}\Psl \Gamma^{\f}\lambda^{\conj}+\frac{1}{16}e^{W}\Gamma^{\f BC}S_{B}\Psi^{\conj}_{C} &=\mathcal{R}_{2\,\f}^{(+)} \otimes \varepsilon_{+}e^{\frac{3i}{2}\chi}\otimes  u_{+}
\nonumber\\
 &
 \,+ \mathcal{R}_{2\, \f}^{(-)} \otimes \varepsilon_{-}e^{-\frac{3i}{2}\chi}\otimes  u_{+}
\end{align}

\noindent obtaining 
\begin{align}
 \mathcal{R}_{2}^{(\pm)a}
={}&
\pm \frac{1}{2} \Psl\gamma^a\lambda^{(\pm)\conj}
\pm \frac18i\cg{3}{ebc}{}
\left(\delta_e^d\gamma^{a}\gamma_{bc}-\delta_e^a\gamma^{d}\gamma_{bc}-\frac13 \gamma^{ad}\gamma_{ebc}
\right)\psi^{(\mp)\conj}_d
\nonumber\\
&
\pm \frac14\cg{2}{eb}{}
\left(\delta_e^d\gamma^{a}\gamma_{b}-\delta_e^a\gamma^d\gamma_b-\frac12\gamma^{ad}\gamma_{eb}
\right)\psi^{(\mp)\conj}_d
\nonumber\\
&
- \frac12\cg{1}{}{e}\gamma^{eda}\psi^{(\mp)\conj}_d
- i\cnpm{1}{}{e}\gamma^{eda}\psi^{(\pm)\conj}_d
+\cnpm{0}{}{}\gamma^{ab}\psi^{(\pm)\conj}_b
\nonumber\\
&
\pm\frac{1}{24}\cg{3}{}{ebc}\gamma^{aebc}\left(\varphi^{(\mp)\conj}+4\rho^{(\mp)\conj}\right)
\mp \frac18i\cg{2}{}{eb}\gamma^{aeb}\left(\varphi^{(\mp)\conj}-4\rho^{(\mp)\conj}\right)
\nonumber\\
&
\pm \frac14i\gamma^a\cgsl{2}\varphi^{(\mp)\conj}
- \frac12i\cg{1}{}{b}
\gamma^{ba}
\varphi^{(\mp)\conj}
+ i\gamma^a\cgsl{1}\rho^{(\mp)\conj}
 \nonumber\\
&
+ \cnpm{1}{}{b}
\gamma^{ba}
\varphi^{(\pm)\conj}
- 2 \gamma^a\cnpmsl{1}\rho^{(\pm)\conj}
- 2ie\gamma^a \cnpm{0}{}{}\rho^{(\pm)\conj}\, ,
\end{align}

\begin{align}
\mathcal{R}^{(\pm)}_{2\, base}
={}&
 \pm 2i\Psl\lambda^{(\pm)\conj}+ 2i\cnpmsl{1}\left(\varphi^{(\pm)\conj}+2\rho^{(\pm)\conj}\right)
 - 2\cnpm{0}{}{}\left(\varphi^{(\pm)\conj}+2\rho^{(\pm)\conj}\right)
\nonumber\\
&
+ \cgsl{1}\left(\varphi^{(\mp)\conj}+2\rho^{(\mp)\conj}\right)
\mp i\cgsl{3}\left(\varphi^{(\mp)\conj}+3\rho^{(\mp)\conj}\right)
\nonumber\\
&
-2i\cnpm{0}{}{}\gamma^d\psi^{(\pm)\conj}_d
- 2\cnpmsl{1}\gamma^d\psi^{(\pm)\conj}_d
+ i\cgsl{1}\gamma^d\psi^{(\mp)\conj}_d
\nonumber\\
&
\mp\frac16\cg{3}{}{ebc}\gamma^{debc}\psi^{(\mp)\conj}_d
\pm \frac12i\cg{2}{}{eb}\gamma^{deb}\psi^{(\mp)\conj}_d
\end{align}

\noindent and 
\begin{align}
\mathcal{R}_{2\,\f}^{(\pm)}
={}&
 \pm i\frac{1}{2} \Psl\lambda^{(\pm)\conj}
 \mp\frac{1}{24}\cg{3}{}{ebc}\gamma^{debc}
\psi^{(\mp)\conj}_d
\mp \frac14i\cg{2}{db}{}\gamma_{b}\psi^{(\mp)\conj}_d
+ \frac12i\cg{1}{}{e}
\gamma^{ed}
\psi^{(\mp)\conj}_d
 \nonumber\\
 &
- \cnpm{1}{}{e}
\gamma^{ed}
\psi^{(\pm)\conj}_d
\mp i\cgsl{3}\rho^{(\mp)\conj}
+ \cgsl{1}\rho^{(\mp)\conj}
+ 2i\cnpmsl{1}\rho^{(\pm)\conj}
- 2\cnpm{0}{}{}\rho^{(\pm)\conj}\, .
\end{align}

In terms of the quantities computed above, the 10-$d$ gravitino equation reduces to the following set of equations for the five-dimensional fields:
\begin{align}
0&= \mathcal{L}^{(\pm)a} - \mathcal{R}_{1}^{(\pm)a} - \mathcal{R}_{2}^{(\pm)a}
\\
0&= \mathcal{L}_{base}^{(\pm)} - \mathcal{R}_{1\, base}^{(\pm)} - \mathcal{R}_{2\, base}^{(\pm)}
\\
0&= \mathcal{L}_{\f}^{(\pm)} - \mathcal{R}_{1\, \f}^{(\pm)} - \mathcal{R}_{2\, \f}^{(\pm)}\, .
\end{align}

Instead of working with the equations of motion given in this form, it is convenient to rewrite them in terms of the fields \eqref{canonical fields 1}-\eqref{canonical fields 4} whose kinetic terms are diagonal. We do so below.

\subsection{Equations of motion in terms of diagonal fields}
The $d=5$ equations of motion for the diagonal fields \eqref{canonical fields 1}-\eqref{canonical fields 4} are given by
\begin{align}
0&=\mathcal{L}_{\tilde{\lambda}}^{(\pm)}-\mathcal{R}_{1\tilde{\lambda}}^{(\pm)}-\mathcal{R}_{2\tilde{\lambda}}^{(\pm)}
\\
0&= \mathcal{L}_{\zeta}^{(\pm)a} - \mathcal{R}_{1\, \zeta}^{(\pm)a} - \mathcal{R}_{2\, \zeta}^{(\pm)a}
\\
0&= \mathcal{L}_{\eta}^{(\pm)} - \mathcal{R}_{1\, \eta}^{(\pm)} - \mathcal{R}_{2\, \eta}^{(\pm)}
\\
0&= \mathcal{L}_{\xi}^{(\pm)} - \mathcal{R}_{1\, \xi}^{(\pm)} - \mathcal{R}_{2\, \xi}^{(\pm)}
\end{align}

\noindent Here,
\begin{align}
\mathcal{L}_{\tilde{\lambda}}^{(\pm)} ={}& e^{W/2}\mathcal{L}_{\lambda}^{(\pm)}
\\
={}&
\Dsl\tilde{\lambda}^{(\pm)}
+\frac14i\Sigma^{-2}\Fsl_{2}\tilde{\lambda}^{(\pm)}
\mp \left(e^{-4U}\Sigma^{-1}+\frac32\Sigma^2\right)\tilde{\lambda}^{(\pm)}
 +\frac34 i e^{\phi} (\pasl a)\tilde{\lambda}^{(\pm)}
\end{align}

\noindent where now $\Dsl \tilde{\lambda}^{(\pm)}=\left(\nablasl \mp\frac{3i}{2}\Asl\right)\tilde{\lambda}^{(\pm)}$ and 
\begin{align}
\mathcal{R}_{1\tilde{\lambda}}^{(\pm)}
={}&
e^{W/2}\mathcal{R}_{1\lambda}^{(\pm)}\\
={}&
 \left(e^{Z+4W}
 -\frac12i e^{-4U} \slashed{K}_1
\mp\frac12i\Sigma\slashed{K}_2\right)\tilde{\lambda}^{(\pm)} 
\mp \Sigma \slashed{L}^{(\pm)}_2 \tilde{\lambda}^{(\mp)}
\nonumber\\
& 
+\left(-\frac14i\gamma^a\cgsl{3}
-  \frac14\gamma^a\cgsl{2}
\pm\frac12\gamma^a\cgsl{1}
\right)\zeta^{(\pm)}_{a}
\mp \left(i\gamma^a \cnpmsl{1}
+ \gamma^a\cnpm{0}{}{}\right)\zeta^{(\mp)}_{a} 
\nonumber\\
& 
+\left(\frac16\cgsl{3}
-  \frac16i\cgsl{2}
\right)\eta^{(\pm)}
\mp \frac{4}{3}i \cnpm{0}{}{}\eta^{(\mp)}
+\frac14\Bigl(\cgsl{3}+i\cgsl{2}\mp 2i\cgsl{1} \Bigr)\xi^{(\pm)}
\nonumber\\
&
\mp \left(\cnpmsl{1}+i \cnpm{0}{}{}\right)\xi^{(\mp)}
\end{align}
Similarly, 
\begin{align}
\mathcal{R}_{2\tilde{\lambda}}^{(\pm)}
={}&
e^{W/2}\mathcal{R}_{2\lambda}^{(\pm)}=
 \pm \gamma^a \Psl {\zeta}^{(\mp)\conj}_{a}\, .
\end{align}

\noindent In the same way, for the $\zeta^{(\pm)}_{a}$ equation of motion we find
\begin{align}
\mathcal{L}_{\zeta}^{(\pm)a} 
={}& e^{W/2}\mathcal{L}^{(\pm)a}
\\
={}& 
\gamma^{abc}\left[D_b +\frac14ie^{\phi} (\pa_b a)\right]\zeta^{(\pm)}_{c} \mp\left(e^{-4U}\Sigma^{-1}+\frac32\Sigma^2\right)\gamma^{ac}\zeta^{(\pm)}_{c}
\nonumber
\\
&
-\frac14i\Sigma^{-2}\gamma^{[c}\Fsl_{2}\gamma^{a]}\zeta^{(\pm)}_{c} 
+\Bigl[i(\pasl U)\gamma^{a}
 \mp ie^{-4U}\Sigma^{-1}\gamma^{a}\Bigr]\xi^{(\pm)}
\nonumber
\\
&
-\frac12i(\Sigma^{-1}\pasl\Sigma)\gamma^{a}\eta^{(\pm)}
+\frac16\Sigma^{-2}\Fsl_{2}\gamma^{a}\eta^{(\pm)}
\pm\frac{i}{3}\left(e^{-4U}\Sigma^{-1}-3\Sigma^2\right)\gamma^{a}\eta^{(\pm)}
\end{align}
\begin{align}
\mathcal{R}_{1\, \zeta}^{(\pm)a}
={}& e^{W/2}\mathcal{R}_{1}^{(\pm)a}
\\
={}& 
\left(-\frac18i\cgtsl{3}
-\frac18 \cgtsl{2}
 \pm\frac14 \cgtsl{1}
\right)\gamma^a\tilde{\lambda}^{(\pm)}
 \mp \left(\frac12i \cntpmsl{1}
+\frac12 \cntpm{0}{}{}\right)\gamma^a\tilde{\lambda}^{(\mp)}
\nonumber\\
&
+\left(e^{Z+4W} \gamma^{ca}
- \frac12i e^{-4U} \gamma^{[c}\slashed K_1\gamma^{a]}
 \mp\frac12i\Sigma\gamma^{[c}\slashed{K}_2\gamma^{a]}\right)\zeta^{(\pm)}_{c}
 \mp\Sigma\gamma^{[c}\slashed{L}_2^{(\pm)}\gamma^{a]} \zeta^{(\mp)}_{c} 
\nonumber\\
&
+\Bigl(-  ie^{Z+4W} 
+\frac12  e^{-4U} \slashed K_1\Bigr)\gamma^{a}\xi^{(\pm)}
+\left(-\frac{2i}{3}e^{Z+4W}\mp\frac16\Sigma\slashed{K}_2\right)\gamma^a\eta^{(\pm)}
\nonumber\\
&
\pm\frac13i\Sigma\slashed{L}_2^{(\pm)}\gamma^a\eta^{(\mp)}
\end{align}

\noindent and
\begin{align}
\mathcal{R}_{2\, \zeta}^{(\pm)a}
={}& e^{W/2}\mathcal{R}_{2}^{(\pm)a}
\\
={}& 
\mp \frac12\Psl\gamma^a\tilde{\lambda}^{(\mp)\conj}
\pm\frac18i\cg{3}{ebc}{}
\left[\frac13\gamma^{da}\gamma_{ebc}+(\delta_e^d\gamma^a-\delta_e^a\gamma^d)\gamma_{bc}
\right] \zeta^{(\mp)\conj}_{d} 
\nonumber\\
&
\pm\left(\frac18\cg{2}{eb}{}\gamma_e\gamma^{da}\gamma_{b}  \mp \frac12\cg{1}{}{b}\gamma^{abd}\right)\zeta^{(\mp)\conj}_{d} 
+\left(- i\cnpm{1}{}{b}\gamma^{abd} 
+ \cnpm{0}{}{}\gamma^{ad}\right)\zeta^{(\pm)\conj}_{d}
\nonumber\\
&
\mp\left(\frac{1}{12}\cgsl{3}
+\frac{i}{12} \cgsl{2} 
\right)\gamma^{a}\eta^{(\mp)\conj}
 + \frac{2}{3}i\cnpm{0}{}{}\gamma^{a}\eta^{(\pm)\conj}
\nonumber\\
&  
\mp \frac18\left(\cgsl{3} -i\cgsl{2}\mp 2i\cgsl{1}\right)\gamma^{a}\xi^{(\mp)\conj}
- \frac12\left( \cnpmsl{1}-i\cnpm{0}{}{}\right)\gamma^a\xi^{(\pm)\conj}\, .
\end{align}

For the $\eta^{(\pm)}$ equation of motion we have
\begin{align}
\mathcal{L}_{\eta}^{(\pm)} 
={}& e^{W/2}\left(\mathcal{L}_{\f}^{(\pm)} + \frac{i}{3}\gamma_{a}\mathcal{L}^{(\pm)a}\right)
\\
={}& 
\frac{2}{3}\left[\Dsl +\frac14ie^{\phi} (\pasl a)\right]\eta^{(\pm)} +\left(-\frac{5}{18}i\Sigma^{-2}\Fsl_{2}
\mp\frac{2}{9}e^{-4U}\Sigma^{-1}
\pm\frac{5}{3}\Sigma^2\right)\eta^{(\pm)}
\nonumber\\
&
+\left[\frac{1}{3}\Sigma^{-2}\gamma^{c}\Fsl_{2}
+i\gamma^c(\Sigma^{-1}\pasl\Sigma) \pm\frac23ie^{-4U}\Sigma^{-1}\gamma^{c}
\mp 2i\Sigma^2\gamma^{c}\right]\zeta^{(\pm)}_{c}
\nonumber\\
&
 \mp\frac{4}{3}e^{-4U}\Sigma^{-1}\xi^{(\pm)}
\end{align}
\begin{align}
\mathcal{R}_{1\eta}^{(\pm)} 
={}& 
e^{W/2}\left(\mathcal{R}_{1\, \f}^{\pm} + \frac{i}{3}\gamma_{a}\mathcal{R}_{1}^{(\pm)a}\right)
\\
={}& 
\left(\frac16\cgtsl{3}
-\frac16i  \cgtsl{2}
\right)\tilde{\lambda}^{(\pm)} 
\mp \frac43i \cntpm{0}{}{}\tilde{\lambda}^{(\mp)}
\nonumber\\
&
+\left(-\frac43ie^{Z+4W} \gamma^{c} 
\mp \frac{1}{3}\Sigma \gamma^c \slashed{K}_2\right)\zeta^{(\pm)}_{c}
\pm \frac{2}{3}i\Sigma \gamma^c \slashed{L}^{(\pm)}_2\zeta^{(\mp)}_{c}
+ \frac{8}{3}e^{Z+4W} \xi^{(\pm)}
\nonumber\\
&
+\left(\frac{10}{9}e^{Z+4W}+\frac13i e^{-4U}\slashed{K}_1
    \pm\frac{1}{9}i\Sigma\slashed{K}_2\right)\eta^{(\pm)} 
   \pm \frac29\Sigma\slashed{L}^{(\pm)}_2\eta^{(\mp)}
\end{align}

\noindent and
\begin{align}
\mathcal{R}_{2\eta}^{(\pm)} 
={}& 
e^{W/2}\left(\mathcal{R}_{2\, \f}^{(\pm)} + \frac{i}{3}\gamma_{a}\mathcal{R}_{2}^{(\pm)a}\right)
\\
={}& 
\mp\frac{1}{6}\gamma^d\left(\cgsl{3} +i\cgsl{2}\right)\zeta^{(\mp)\conj}_{d}
+ \frac43i\cnpm{0}{}{}\gamma^{d}\zeta^{(\pm)\conj}_{d}
\nonumber\\
&
\pm \frac{1}{18}\left(i\cgsl{3}-\cgsl{2}\mp 6\cgsl{1}\right)\eta^{(\mp)\conj}
- \left( \frac{10}{9}\cnpm{0}{}{}
+\frac{2i}{3}\cnpmsl{1}\right)\eta^{(\pm)\conj} 
\nonumber\\
& 
\mp\frac16\left( i\cgsl{3} +\cgsl{2}\right)\xi^{(\mp)\conj}
-\frac{4}{3}\cnpm{0}{}{}\xi^{(\pm)\conj}\, .
\end{align}

\noindent Finally, for the $\xi^{(\pm)}$ equation of motion we have
\begin{align}
\mathcal{L}_{\xi}^{(\pm)} 
={}& e^{W/2}\left( i\gamma_{a}\mathcal{L}^{(\pm)a}
+\mathcal{L}^{(\pm)}_{base}-\mathcal{L}_{\f}^{(\pm)}\right)
\\
={}& 
 2\left[\Dsl+\frac14ie^{\phi} (\pasl a)\right]\xi^{(\pm)}
+\frac12i\Sigma^{-2}\Fsl_{2}\xi^{(\pm)}
\pm 3\left(2e^{-4U}\Sigma^{-1}
-\Sigma^{2}\right)\xi^{(\pm)}
\nonumber\\
&
+\Bigl[\mp 4ie^{-4U}\Sigma^{-1}\gamma^{c}
-4i\gamma^{c}(\pasl U)\Bigr]\zeta^{(\pm)}_{c} 
\mp\frac{8}{3}e^{-4U}\Sigma^{-1}\eta^{(\pm)}\\
\nonumber\\
\mathcal{R}_{1\xi}^{(\pm)} 
={}& 
e^{W/2}\left(i\gamma_{a}\mathcal{R}_{1}^{(\pm)a}
+ \mathcal{R}_{1\, base}^{(\pm)}-\mathcal{R}_{1\, \f}^{(\pm)} \right)
\\
={}& 
\left(\frac12\cgtsl{3}
+\frac12i  \cgtsl{2}
\mp i\cgsl{1}
\right)\tilde{\lambda}^{(\pm)} 
\mp\left(2\cntpmsl{1}
+2i\cntpm{0}{}{}\right)\tilde{\lambda}^{(\mp)}
\nonumber\\
&
+\left(-4ie^{Z+4W} \gamma^{c} 
+ 2e^{-4U}\gamma^{c}\slashed{K}_1\right)\zeta^{(\pm)}_{c} 
+  \frac{16}{3}e^{Z+4W}\eta^{(\pm)}
\nonumber\\
&
+\left( 6e^{Z+4W} 
-3i  e^{-4U} \slashed K_1
 \pm i\Sigma \slashed{K}_{2}\right)\xi^{(\pm)}
\pm 2 \Sigma\slashed{L}^{(\pm)}_{2}\xi^{(\mp)}
\end{align}

\noindent and
\begin{align}
\mathcal{R}_{2\xi}^{(\pm)} 
={}& 
e^{W/2}\left(i\gamma_{a}\mathcal{R}_{2}^{(\pm)a}
+ \mathcal{R}_{2\, base}^{(\pm)}-\mathcal{R}_{2\, \f}^{(\pm)} \right)
\\
={}& 
\mp\left(\frac{1}{2}\gamma^d\cgsl{3}
- \frac12i\gamma^d\cgsl{2}
\mp i\gamma^d\cgsl{1} 
\right)\zeta^{(\mp)\conj}_{d}
+ \left(2i\cnpm{0}{}{}\gamma^{d}
-2\gamma^d\cnpmsl{1}\right) \zeta^{(\pm)\conj}_{d} 
\nonumber\\
&
\mp\left(\frac{1}{3}i\cgsl{3}
+\frac{1}{3} \cgsl{2}
\right)\eta^{(\mp)\conj}
-\frac83\cnpm{0}{}{}\eta^{(\pm)\conj}
\mp\frac34\cgsl{2}{\xi}^{(\mp)\conj}\, .
\end{align}

\newpage

\providecommand{\href}[2]{#2}\begingroup\raggedright\endgroup


\end{document}